\documentclass[aps,prl,twocolumn,superscriptaddress, 10pt]{revtex4-1}
\usepackage{graphics,graphicx, array, afterpage}
\usepackage{qcircuit,amsmath}
\usepackage{amsfonts}
\usepackage{grffile}
\usepackage[export]{adjustbox}
\usepackage{lineno}
\usepackage{longtable}
\usepackage{array}
\usepackage{braket}
\usepackage{multirow}

\newcommand{\Gate}[1]{\textsc{#1}}
\newcommand{\cnotgate}{\Gate{cnot}} 
\newcommand{\xxgate}{\Gate{xx}} 
\newcommand{\rgate}{\Gate{r}} 
\newcommand{\Yb}{$^{171}{\rm{Yb}}^{+} $}

\begin{document}

\title{Parallel Entangling Operations on a Universal Ion Trap Quantum Computer}

\author{C. Figgatt}
\affiliation{Joint Quantum Institute, Department of Physics, and Joint Center for Quantum Information and Computer Science, University of Maryland, College Park, MD 20742, USA}
\author{A. Ostrander}
\affiliation{Department of Physics and Joint Center for Quantum Information and Computer Science, University of Maryland, College Park, MD 20742, USA}
\author{N. M. Linke}
\author{K. A. Landsman}
\author{D. Zhu}
\affiliation{Joint Quantum Institute, Department of Physics, and Joint Center for Quantum Information and Computer Science, University of Maryland, College Park, MD 20742, USA}
\author{D. Maslov}
\affiliation{National Science Foundation, Alexandria, VA 22314, USA}
\affiliation{Joint Quantum Institute, Department of Physics, and Joint Center for Quantum Information and Computer Science, University of Maryland, College Park, MD 20742, USA}
\author{C. Monroe}
\affiliation{Joint Quantum Institute, Department of Physics, and Joint Center for Quantum Information and Computer Science, University of Maryland, College Park, MD 20742, USA}
\affiliation{IonQ Inc., College Park, MD 20742, USA}

\date{\today}


\maketitle 

The circuit model of a quantum computer consists of sequences of gate operations between quantum bits (qubits), drawn from a universal family of discrete operations \cite{NielsenChuang11}. 
The ability to execute parallel entangling quantum gates offers clear efficiency gains in numerous quantum circuits \cite{CleveQFT00, MaslovQFTStabilizerParallel07, MaslovAdder08} as well as for entire algorithms such as Shor's factoring algorithm \cite{FowlerShorParallel04} and quantum simulations \cite{NamQSim18,KivlichanQChemParallel17}.  In cases such as full adders and multiple-control Toffoli gates, parallelism can provide an exponential improvement in overall execution time. 
More importantly, quantum gate parallelism is essential for the practical fault-tolerant error correction of qubits that suffer from idle errors \cite{Steane1999,AharonovDecohere96}. 
The implementation of parallel quantum gates is complicated by potential crosstalk, especially between qubits fully connected by a common-mode bus, such as in Coulomb-coupled trapped atomic ions \cite{WinelandBlatt2008, MonroeScaling13} or cavity-coupled superconducting transmons \cite{DevoretSCscalability13}. 
Here, we present the first experimental results for parallel 2-qubit entangling gates in an array of fully-connected trapped ion qubits. 
We demonstrate an application of this capability by performing a 1-bit full addition operation on a quantum computer using a depth-4 quantum circuit 
\cite{FeynmanQuantum85, MaslovAdder08, DraperAdder06}. 
These results exploit the power of highly connected qubit systems through classical control techniques, and provide an advance toward speeding up quantum circuits and achieving fault tolerance with trapped ion quantum computers.

Trapped atomic ions are among the most advanced qubit platforms \cite{WinelandBlatt2008, MonroeScaling13}, with atomic clock precision and the ability to perform gates in a fully-connected and reconfigurable qubit network \cite{Debnath16}. The high connectivity between trapped ion qubits \cite{LinkeComparison17} is mediated by optical forces on their collective motion \cite{CiracZoller95, Molmer99, Solano99, Milburn00}, 
and can be scaled in a modular fashion using a variety of methods \cite{WinelandBlatt2008, MonroeScaling13}. 
Within a single large chain of ions, gates can be performed by appropriately shaping the laser pulses that drive select ions within the chain. 
Here, the target qubits become entangled through their Coulomb-coupled motion, and the laser pulse is modulated such that the motional modes are disentangled from the qubits at the end of the operation \cite{Zhu06PRL, ZhuGates06, Choi14}. 
The execution of multiple parallel gates in this way requires more complex pulse shapes, not only to disentangle the motion but also to entangle exclusively the intended qubit pairs.  We achieve this type of parallel operation by designing appropriate optical pulses using nonlinear optimization techniques.

We perform parallel gates on a chain of five atomic \Yb ions, with resonant laser radiation used to laser-cool, initialize, and measure the qubits. 
Coherent quantum gate operations are achieved by applying counterpropagating Raman beams from a single mode-locked laser, which form beat notes near the qubit difference frequency. 
Single-qubit gates are generated by tuning the Raman beatnote to the qubit frequency splitting $\omega_0$ and driving resonant Rabi rotations ($\rgate$ gates) of defined phase and duration. 
Two-qubit ($\xxgate$) gates are realized by illuminating two ions with beams that have beat-note frequencies near the motional sidebands, creating an effective Ising interaction between the ions via transient entanglement through the modes of motion \cite{Molmer99, Solano99, Milburn00}. We use an amplitude-modulated pulse-shaping scheme that provides high-fidelity entangling gates on any ion pair \cite{Debnath16, ZhuGates06,Choi14}; frequency \cite{LeungFMGates18} or phase \cite{GreenPhaseModulated15} modulation of the laser pulses would also suffice. 
(See Methods for additional experimental setup details.)


In order to perform parallel entangling operations involving $M$ independent pairs of qubits in a chain of $N \geq 2M$ ions with $N$ motional modes at frequencies $\omega_k$, a shaped qubit-state-dependent force is applied to the involved ions using bichromatic beat notes at $\omega_0 \pm \mu$, resulting in the evolution operator \cite{ZhuGates06,Zhu06PRL}
\begin{align} \label{eqn:ParallelUnitary}
U_{||}(\tau) &= \exp\left( \sum_{i=0}^{2M} \hat{\phi}_i(\tau) \sigma_i^x + i \sum_{i<j}^{2M} \chi_{ij}(\tau) \sigma_i^x \sigma_j^x \right),
\end{align}
where $\tau$ is the gate time. 
The first operator describes state-dependent displacements of each mode $k$ in phase space \cite{ZhuGates06}, with 
$\hat{\phi}_i(\tau) = \sum_k \left( \alpha_{i,k}(\tau) \hat{a}_k^{\dag} - \alpha_{i,k}^*(\tau) \hat{a}_k \right)$  and accumulated displacement value
\begin{equation}\label{eqn:alpha}
\alpha_{i,k}(\tau) = \int_0^{\tau} \eta_{i,k} \Omega_i(t) \sin(\mu t) e^{i\omega_k t} dt.
\end{equation}
Here, $\hat{a}_k^{\dag} $ and $\hat{a}_k$ are the raising and lowering operators for mode $k$, 
$\eta_{i,k}$ is the Lamb-Dicke parameter coupling qubit $i$ to mode $k$, and 
$\Omega_i(t)$ is the Rabi frequency of the $i$th ion, proportional to the amplitude-modulated laser intensity hitting the ion. 
In order to generate independent $\xxgate$ gates, we implement separate control signals for each of the $M$ ion pairs we want to entangle, thereby providing enough parameters to simultaneously entangle only the desired ion pairs.
The parameter $\chi_{ij}$ in Equation \ref{eqn:ParallelUnitary} entangles qubits $i$ and $j$ and is given by 
\begin{align}\label{eqn:chi}
\chi_{ij}(\tau) &= 2 \int_0^{\tau} dt' \int_0^{t'} dt \sum_k \eta_{i,k} \eta_{j,k} \Omega_i(t) \Omega_j(t) \times \nonumber\\
& \qquad \sin(\mu t) \sin(\mu t') \sin(\omega_k (t'-t)) .
\end{align}

At the end of the gate operation, the $2MN$ accumulated displacement values in Equation \ref{eqn:alpha} (for the $2M$ involved ions and $N$ modes) should vanish so that all mode trajectories close in phase space and there is no residual qubit-motion entanglement. For each of the $M$ desired entangled pairs, we require $\chi_{ij} = \pi/4$ for maximal entanglement (or other nonzero values for partial entanglement), and for the other crosstalk pairs of qubits, $\chi_{ij} = 0$.
This yields a total of $2MN+\binom{2M}{2}$ constraints for designing appropriate pulse sequences 
$\Omega_i(t)$ to implement the $M$ parallel entangling gates.
To provide optimal control during the gate and fulfill these constraints, we divide the laser pulse at ion $i$ into $S$ segments of equal time duration $\tau/S$, and vary the amplitude in each segment as an independent variable.

While the $2MN$ motional mode constraints (Equation \ref{eqn:alpha}) are linear in the control parameters $\Omega_i(t)$, the $\binom{2M}{2}$ entanglement constraints (Equation \ref{eqn:chi}) are quadratic. Finding pulse solutions to this non-convex quadratically constrained quadratic program (QCQP) is an NP-hard problem in general. Because analytical approaches are intractable, we use numerical optimization techniques to find solutions. Further discussion of the constraint problem setup and derivation of the fidelity of simultaneous $\xxgate$ gate operations as a function of the above control parameters is provided in the Methods and \cite{FiggattThesis18}.


\begin{figure*}
\centering
\begin{tabular}[c]{m{5cm} c}
\multicolumn{1}{l}{\begin{tabular}{@{}c l@{}} \bf{(a)} & {\normalfont ions} \\& $i=(1,4)$ \end{tabular}} &
\multicolumn{1}{c}{\raisebox{22.5mm}{\includegraphics[width=0.9\textwidth,valign=T]{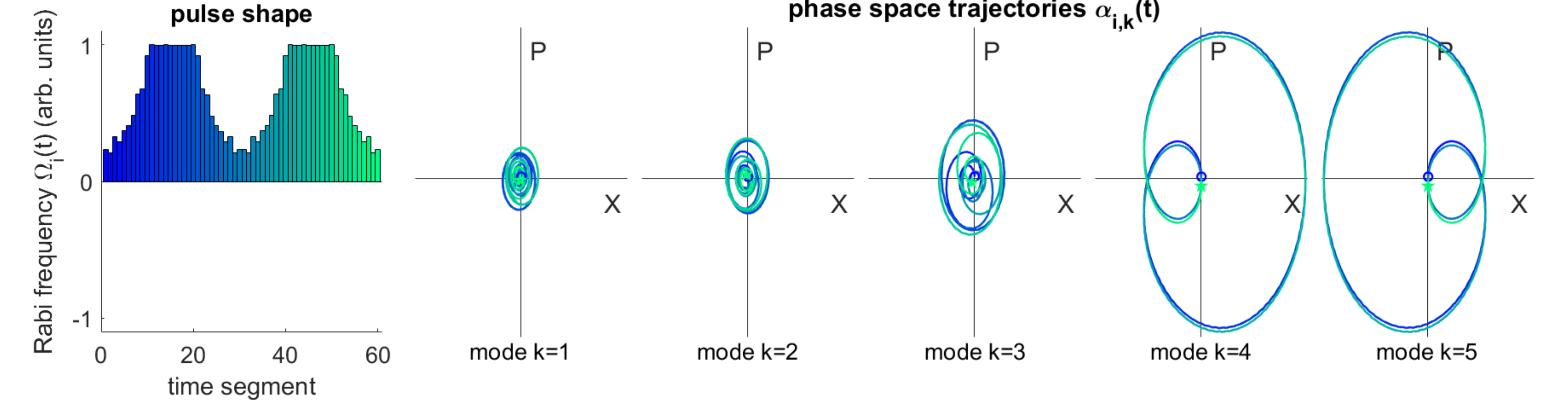}}} \\
\multicolumn{1}{l}{\begin{tabular}{@{}c l@{}} \bf{(b)} & {\normalfont ions} \\& $i=(2,5)$ \end{tabular}} &
\multicolumn{1}{c}{\raisebox{22.5mm}{\includegraphics[width=0.9\textwidth,valign=T]{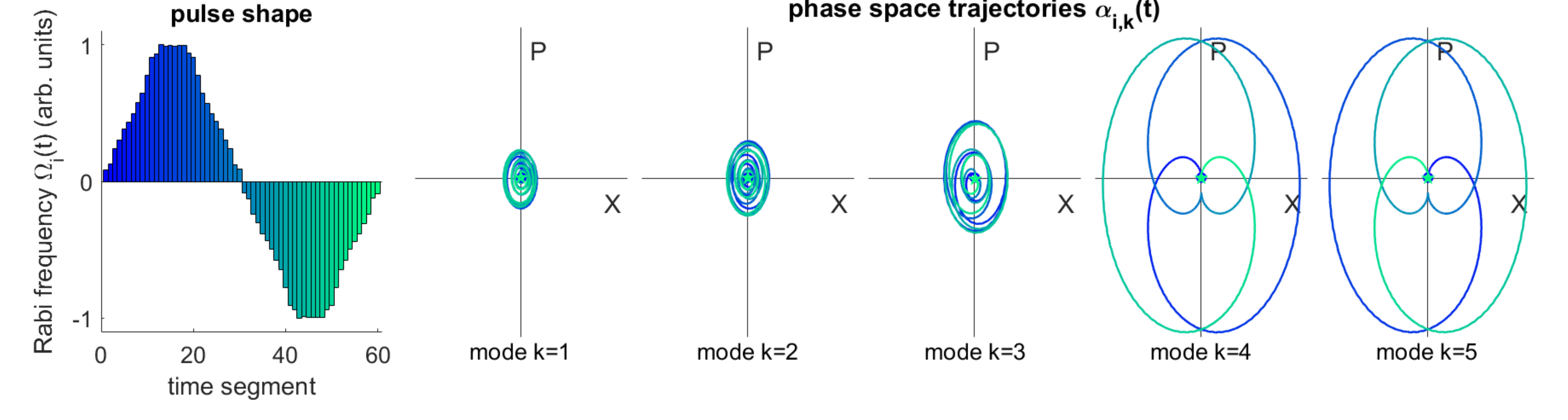}}}\\ 
\end{tabular}
\caption[Pulse shape for (1,4), (2,5).]{
Laser pulse shape solutions (leftmost panels) and theoretical phase space trajectories $\alpha_{i,k}$ for each mode $k$ correlated with ion $i$ (right five panels) for parallel $\xxgate$ gates on (a) ions (1,4) and (b) ions (2,5). The pulse shape solutions are expressed in terms of the time-dependent Rabi frequency $\Omega_i(t)$ experienced by both ions in each pair, and is broken into $S=60$ segments with a total gate time of 250 $\mu$s. Negative Rabi frequencies correspond to an inverted phase of the beatnote. The 5 modes of motion have frequencies $\omega_k/2\pi = \{3.045, 3.027, 3.005, 2.978, 2.946\}$ MHz, and with the constant laser beatnote detuning set to $\mu = 2.962$ MHz, the nearby modes 4 and 5 feel the largest displacements. The phase space trajectories begin at the blue circles and follow the continuous path to the green star, with the color shading of the trajectory corresponding to the pulse shape in time at left. The sum of the normalized area enclosed by all 5 modes is set to $\pi/4$.
}
\label{fig:Theory1425}
\end{figure*}

Parallel gates are designed for 2 independent ion pairs in a 5-ion chain. 
Pulse sequences were designed by solving an optimization problem that took into account laser power and the constraints on the $\alpha$ and $\chi$ parameters. 
Sequences are calculated for a gate time of $\tau_{gate}=250$ $\mu$s, which is comparable to the standard 2-qubit $\xxgate$ gates already used on the experiment \cite{Debnath16}, and for a range of detunings $\mu$. This generated a selection of solutions, which are tested on the experimental setup; the solution generating the highest-quality gate using the least amount of power is chosen. 

Experimental gates are found for 6 ion pair combinations: 
\{(1,4), (2,5)\}; \{(1,2), (3,4)\}; \{(1,5), (2,4)\}; \{(1,4), (2,3)\}; \{(1,3), (2,5)\}; and \{(1,2), (4,5)\}. 
Figure \ref{fig:Theory1425} shows the pulse sequence applied to each entangled pair to construct a set of parallel 2-qubit gates on ions (1,4) and (2,5), as well as the trajectories in phase space of each mode-pair interaction. 
The five transverse motional modes in this 5-ion chain have sideband frequencies $\{\omega_k/2\pi\}=\{3.045, 3.027, 3.005, 2.978, 2.946 \}$ MHz, where mode 1 is the 3.045 MHz common mode. The phase space trajectories show that modes 4 and 5, closest to the selected detuning of $\mu=2.962$ MHz, exhibit the greatest displacement, and contribute the most to the final spin-spin entanglement by enclosing a larger area of phase space. 
Negative-amplitude pulses are implemented by applying a $\pi$ phase shift to the control signal, allowing the entangling pairs to continue accumulating entanglement while cancelling out accumulated entanglement with cross-talk pairs. 
Consequently, all pulse solutions feature similar patterns with symmetric phase flips on one pair to cancel out crosstalk entanglement. 
Pulse shapes and phase space trajectories for the additional solutions are given in \cite{FiggattThesis18}.

\afterpage{\clearpage}


\begin{figure*}
\centering
\begin{tabular}[c]{c c}
\multicolumn{1}{l}{\bf{(a)}} & \multicolumn{1}{l}{\bf{(b)}} \\
\multicolumn{1}{c}{\includegraphics[width=0.5\textwidth,valign=T]{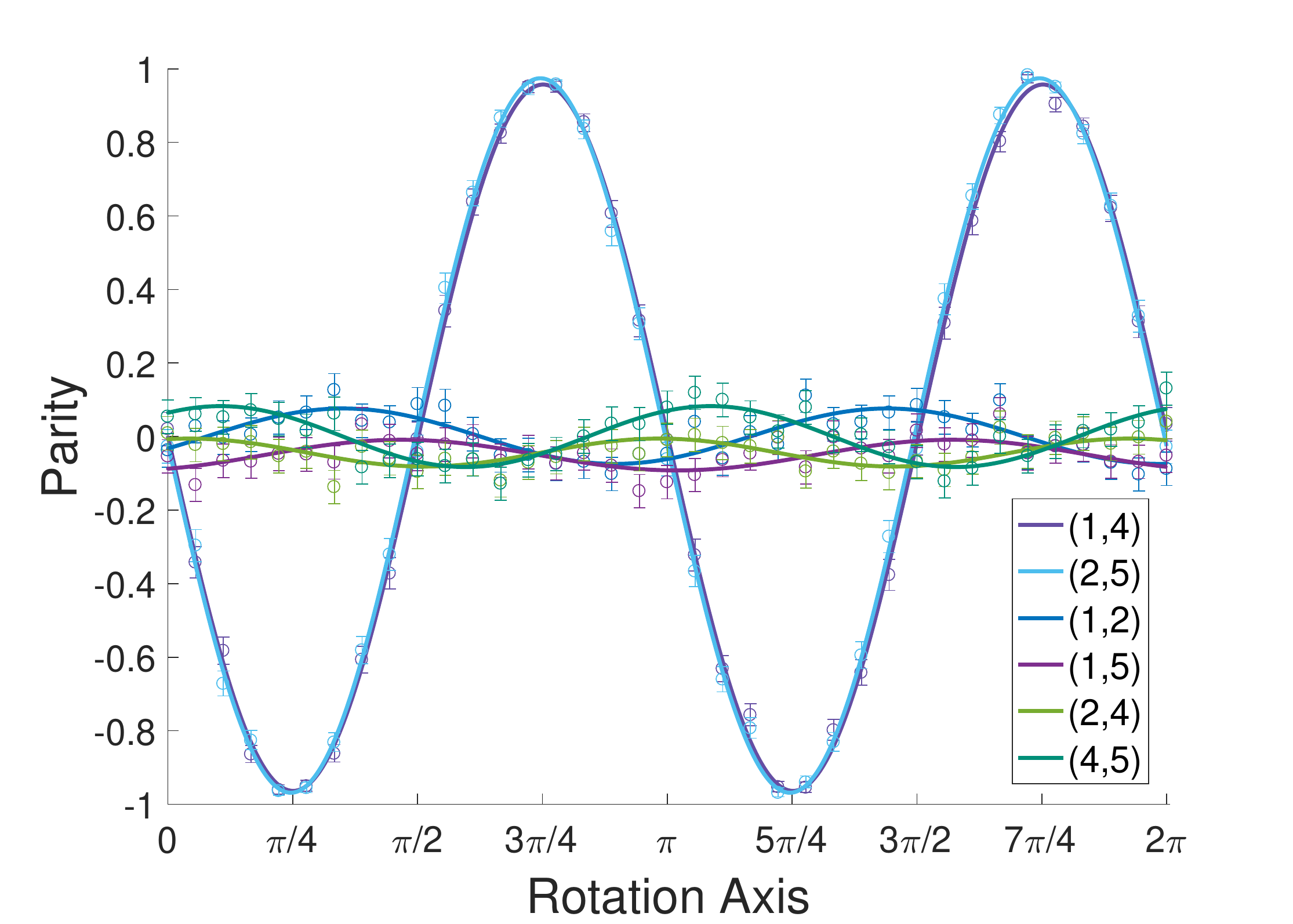}} & 
\multicolumn{1}{c}{\includegraphics[width=0.5\textwidth,valign=T]{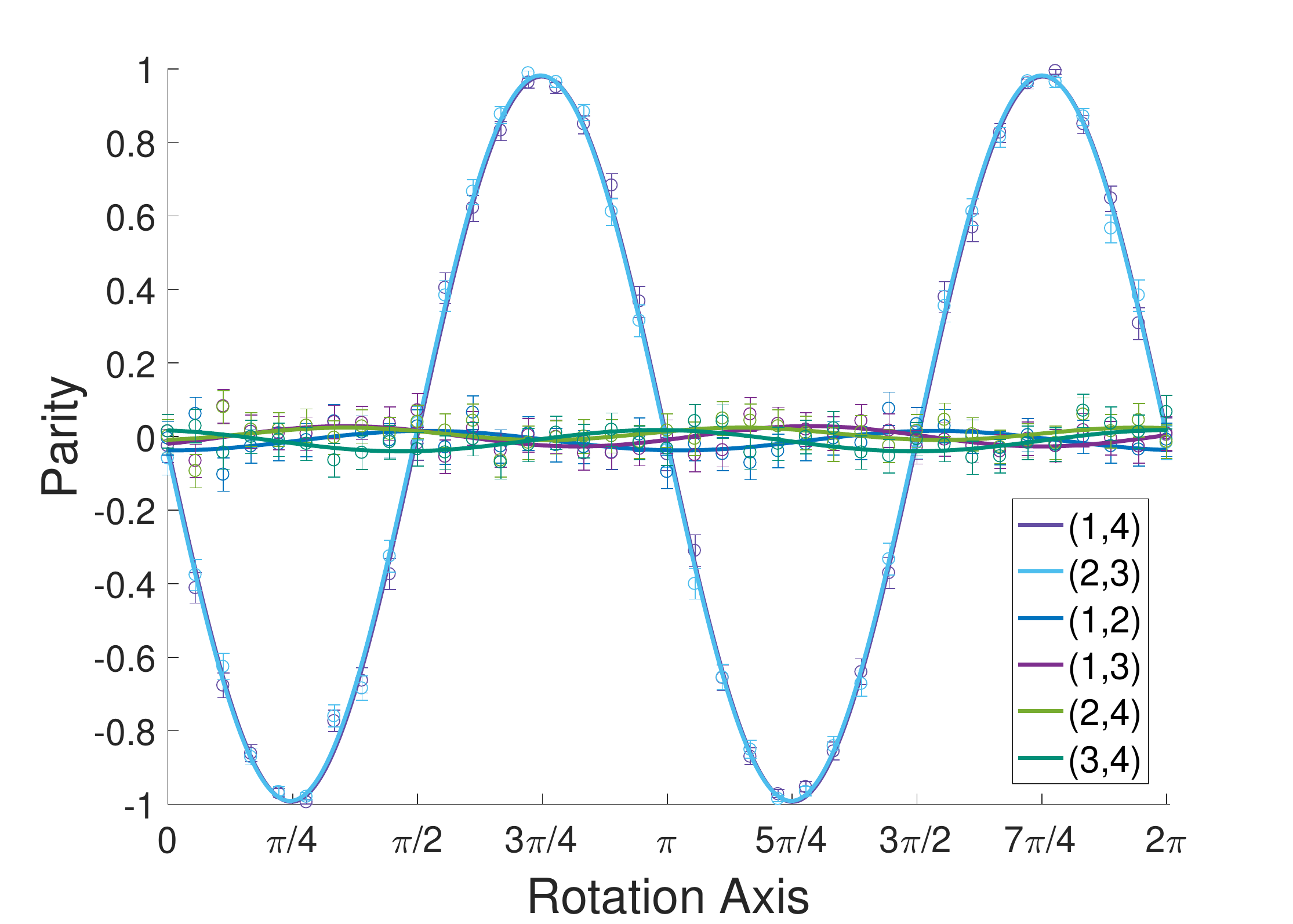}}\\ 
\end{tabular}
\caption[Parity curves for parallel 2-qubit gates on several sets of ions.]{Parity curves and fidelities for parallel $\xxgate$ gates on two example sets of ions. Circles indicate data, with matching-color lines indicating calculated fit. Additional data is shown in the Methods. 
(a) Ions (1,4) and (2,5), yielding fidelities of 96.5(4)\% and 97.8(3)\% on the respective entangled pairs, with an average crosstalk error of 3.6(3)\%. 
(b) Ions (1,4) and (2,3), yielding fidelities of 98.8(3)\% and 99.0(3)\% on the respective entangled pairs, with an average crosstalk error of 1.4(3)\%. 
The quoted errors are statistical.
}
\label{fig:ParityCurves}
\end{figure*}

We characterize the experimental gate fidelities by measuring the selected output qubits in different bases and extracting the parity as a witness operator \cite{Sackett4ParticleEntanglement00, KimYbEntanglement09, ManningThesis14}, as described in the Methods. 
Fitted parity curves are shown in Figure \ref{fig:ParityCurves}. Entangling gate fidelities are typically between 96-99\%, with crosstalk of a few percent. 
Crosstalk errors are estimated by fitting the crosstalk pair populations and parity in the same way as above. Crosstalk fidelities for all pairs were close to 25\%, which indicate a complete statistical mixture. The given errors are statistical. All data has been corrected for state preparation and measurement (SPAM) errors of 3-5\%, as described in \cite{Debnath16,Linke422_17,FiggattThesis18}.


\begin{figure}
\centering
\includegraphics[width=\columnwidth]{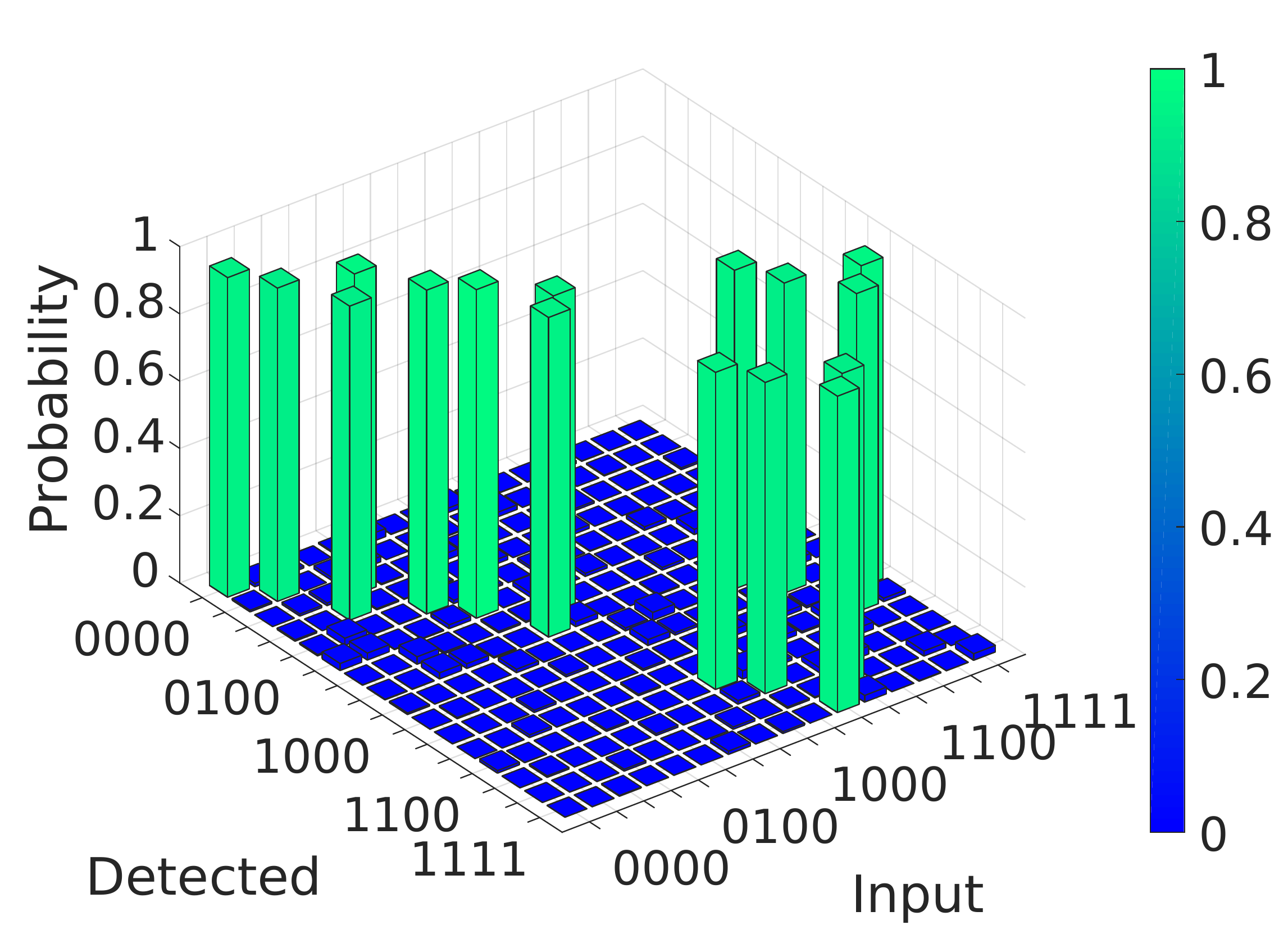}
\caption[Simultaneous $\cnotgate$ gates.]{Data for sumiltaneous $\cnotgate$ gates on ions (1,4) and (2,3), with an average process fidelity of 94.5(2)\%.
}
\label{fig:SCN1423}
\end{figure}

As an example application of a parallel operation useful for error correction codes \cite{MaslovQFTStabilizerParallel07}, we perform a pair of $\cnotgate$ gates in parallel on two pairs of ions. The $\cnotgate$ gate sequence (compiled version with $\rgate$ and $\xxgate$ gates shown in \cite{Debnath16}) is performed simultaneously on the pair $(1,4)$, with ion 1 acting as the control and ion 4 acting as the target, and on the pair $(2,3)$, with ion 2 acting as the control and ion 3 acting as the target. 
The simultaneous $\cnotgate$ gates are performed for each of the 16 possible bitwise inputs, and population data for the 16 possible bitwise outputs is shown in Figure \ref{fig:SCN1423} with an average process fidelity of 94.5(2)\%.


\begin{figure*}
\centering
\begin{tabular}[b]{l l}
\multicolumn{1}{l}{\bf{(a)}} & \multicolumn{1}{l}{\bf{(c)}}\\
&\\
\raisebox{35mm}{
\begin{tabular}[b]{l}
\multicolumn{1}{l}{
\Qcircuit @C=1.4em @R=1.15em @!R {
\lstick{x}		& \ctrl{3}	 & \ctrl{1} 	& \qw	& \qw	& \qw  & \push{x}\\
\lstick{y}		& \ctrl{2} 	& \targ	& \ctrl{2}	& \ctrl{1}	& \qw & \push{y'}\\
\lstick{C_{in}}	& \qw 	& \qw	& \ctrl{1}	& \targ	& \qw & \push{S}\\
\lstick{0} 		& \targ 	& \qw	& \targ	& \qw 	& \qw & \push{C_{out}}
}
}\\ \\ \\
\multicolumn{1}{l}{\bf{(b)}} \\
\\
\multicolumn{1}{l}{
\Qcircuit @C=1.5em @R=.55em @!R {
\lstick{x}		& \qw	 	& \ctrl{1} 	& \ctrl{3}	& \qw	& \qw			& \qw  & \push{x}\\
\lstick{y}		& \ctrl{2} 		& \targ	& \qw	& \ctrl{1}	& \qw			& \qw & \push{y'}\\
\lstick{C_{in}}	& \qw 		& \ctrl{1}	& \qw	& \targ	& \ctrl{1}			& \qw & \push{S}\\
\lstick{0} 		& \gate{V} 	& \gate{V}	& \gate{V}	& \push{\rule{1.0em}{0.04em}}\qw	& \gate{V^{\dag}}	& \qw & \push{C_{out}}
\gategroup{1}{3}{4}{3}{1em}{--}	\gategroup{1}{4}{4}{5}{1em}{--}
}
} \\
\end{tabular}
}
&
\multicolumn{1}{l}{\raisebox{10mm}{\includegraphics[width=250pt]{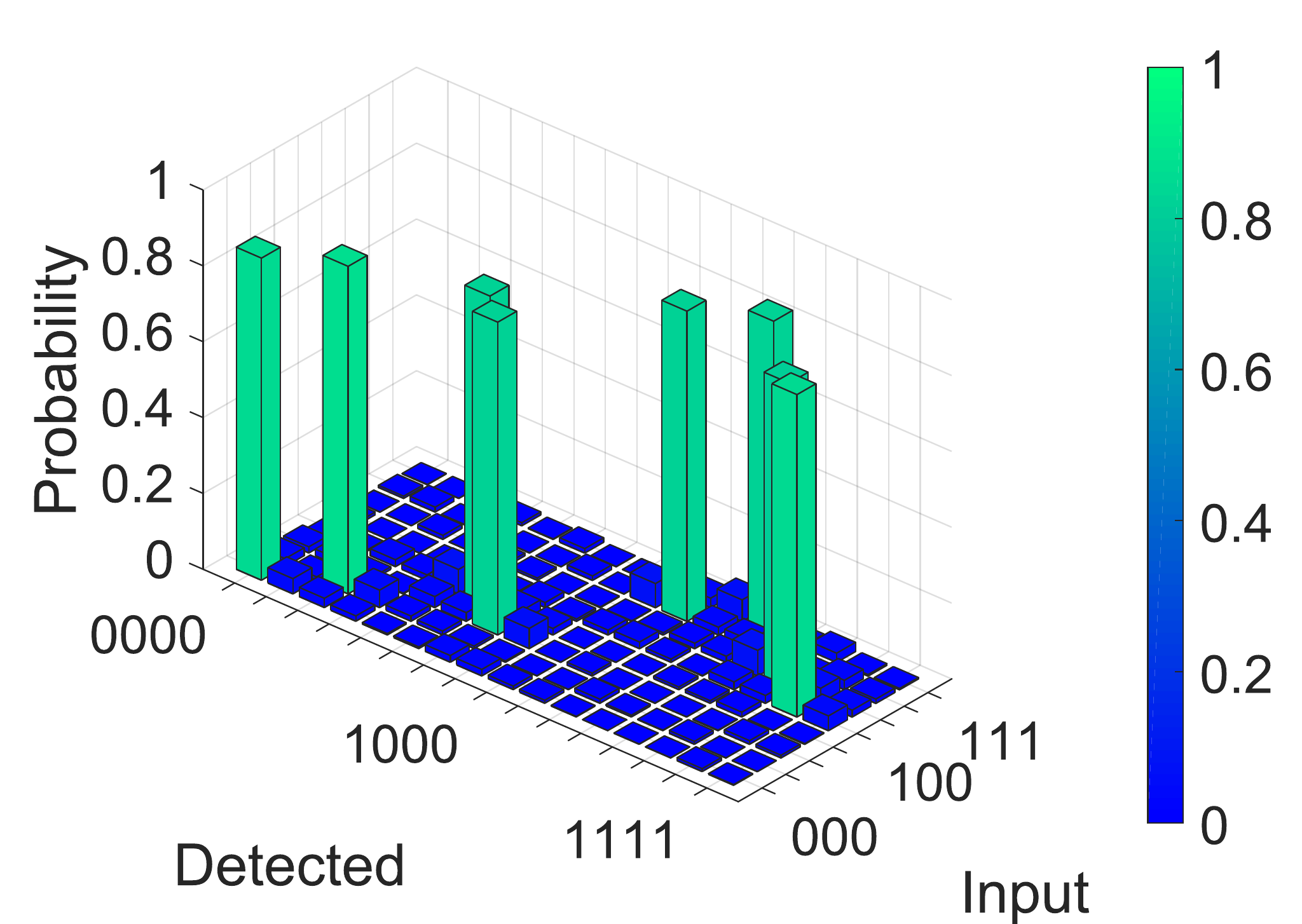}}} \\
\end{tabular}
\caption[Full adder circuits.]{(a) Feynman's original quantum full adder \cite{FeynmanQuantum85}. (b) Optimized full adder with 2-qubit gate depth 4 \cite{MaslovAdder08}. The two parallel 2-qubit operations are outlined in dashed boxes. (c) Data for full adder experimental implementation using simultaneous 2-qubit gates on ions (1,2,4,5), with an average process fidelity of 83.3(3)\%.
}
\label{fig:AdderCircuits}
\end{figure*}

Another application that benefits from the use of parallel entangling operations is the quantum full adder. 
In modern classical computing, a full adder is a basic circuit that can be cascaded to add many-bit numbers, and can be found in processors as components of arithmetic logic units (ALU's) and performing low-level operations such as computing register addresses. 
In quantum computing, adders can be used in a similar fashion to perform arithmetic operations over quantum registers (e.g. \cite{NamQSim18}); some algorithms are dominated by the adders, notably Shor's integer factoring algorithm. 
The quantum full adder requires 4 qubits, 3 for the primary inputs $x$, $y$, and the carry bit $C_{in}$, and the fourth initialized to $\ket{0}$. The four outputs consist of the first input, $x$, simply continuing through; $y'$, which carries $x \oplus y$ (an additional $\cnotgate$ can be added to extract $y$ if desired); and the sum $S$ and output carry $C_{out}$, which together comprise the 2-bit result of summing $x$, $y$, and $C_{in}$, where $C_{out}$ is the most significant bit and hence becomes the carry bit to the next adder in a cascade, and $S$ is the least significant bit. We can also write the sum as $S=x \oplus y \oplus C_{in}$ and the output carry as $C_{out} = (x \cdot y) \oplus (C_{in}\cdot (x \oplus y))$. 
Feynman first designed such a circuit using $\cnotgate$ and Toffoli gates \cite{FeynmanQuantum85}, shown in Figure \ref{fig:AdderCircuits}(a), which would require 12 $\xxgate$ gates to implement on an ion trap quantum computer. 
A more efficient circuit requires at most 6 2-qubit interactions \cite{MaslovAdder08}, and features a gate depth of only 4 if simultaneous 2-qubit operations are available, as shown by the dashed outlines in Figure \ref{fig:AdderCircuits}(b).

The full adder is implemented using 2 different parallel $\xxgate$ gate configurations, as well as the single-qubit rotations and additional $\xxgate$ gates shown in Figure \ref{fig:OptimizedAdderCircuit} in the Methods. The parallel gates require different amounts of entanglement, equivalent to implementing a fully-entangling $\xxgate\left(\chi_{ij}= \frac{\pi}{4} \right)$ gate and a partially-entangling $\xxgate\left( \frac{\pi}{8} \right)$ gate in parallel. This is experimentally implemented by independently adjusting the optical power supplied to each gate; discussions of the calibration independence of these parallel gates and fidelity data for such an operation are given in the Methods. The inputs $x$, $y$, $C_{in}$, and $0$ are mapped to the qubits $(1,2,4,5)$ respectively. 
Figure \ref{fig:AdderCircuits}(c) shows the resulting data from implementing this computation, with all 8 possible bitwise inputs on the 3 input qubits, and displaying the populations in all 16 possible bitwise outputs on the 4 qubits used. The data yields an average process fidelity of 83.3(3)\%.


Faster serial 2-qubit gates can be accomplished with more optical power, but this speedup is limited by sideband resolution, a limitation that gets worse as the processor size grows due to spectral crowding. Parallel 2-qubit operations are a tool to speed up computation that avoids this problem. The control scheme presented here for parallel 2-qubit entangling gates in ions also suggests a method for performing multi-qubit entanglement in a single operation, discussed in the Methods.

When pre-calculating optimal solutions, the number of constraints grows polynomially in the number of ions and entangling pairs. 
Two parallel $\xxgate$ gates in a chain of $N$ ions requires $4N+6 \sim O(N)$ constraints, so the problem size growth is linear in $N$. Entangling more pairs in parallel enlarges the problem size quadratically: entangling $M$ pairs involves the interactions of $2M$ ions, 
yielding 
$\binom{2M}{2} = 2M^2 - M \sim O(M^2)$ 
spin-spin interactions to control and $2MN$ spin-motional entanglements to close. Scaling both the number of entangled pairs $M$ and the number of ions $N$ in the chain therefore gives a total number of constraints of 
$2MN + 2M^2 - M \sim O(M^2+MN)$. 
On very long chains, not all ion-ion connections will be directly available \cite{DuanEqualSpacing09}, reducing the number of quadratic constraints on crosstalk pairs that must be considered, indicating that this is an upper bound on the scaling.

Several lines of future inquiry may help increase the theoretical solution fidelity. 
Easing constraints on the power needed may allow for higher-fidelity solutions to be calculated, although increasing power on the experiment can exacerbate errors due to Raman beam noise.  
Investigating whether the constraint matricies in Equation \ref{eqn:ParallelConstraintsVectors} (see Methods) can be modified to become positive or negative semidefinite may provide improvements, as convex QCQP's are readily solved using semidefinite programming techniques and could allow for higher-fidelity solutions. 
However, these problems are all ones of overhead. Once a high-quality gate solution is implemented on the experiment, no further calculations are needed; only a single calibration is required to compensate for Rabi frequency drifts.

\afterpage{\clearpage}
\newpage
\pagebreak

\twocolumngrid

\textbf{Data availability.} All relevant data is available from the corresponding author upon request.

\afterpage{\clearpage}

\bibliographystyle{apsrev4-1}

%

\begin{acknowledgments}
\textbf{Acknowledgements} We thank S. Wang, Z. Gong, S. Debnath, P. H. Leung, Y. Wu, and L. Duan for helpful discussions.  
Circuits were drawn using the qcircuit.tex package. This work was supported by the ARO with funds from the IARPA LogiQ program, the AFOSR MURI program, and the NSF Physics Frontier Center at JQI. 

This material was partially based on work supported by the National Science Foundation during D.M.'s assignment at the Foundation. Any opinion, finding, and conclusions or recommendations expressed in this material are those of the authors and do not necessarily reflect the views of the National Science Foundation.

\textbf{Author contributions} 
C.F., A.O., N.M.L., K.A.L., D.M., and C.M. designed the research; C.F., N.M.L., K.A.L., D.Z., and C.M. collected and analyzed data; C.F. performed the theory derivations; A.O. performed the pulse sequence optimizations; and C.F., A.O., N.M.L., K.A.L., D.Z., D.M., and C.M. all contributed to the manuscript.

\textbf{Corresponding author}
Correspondence should be addressed to C.F. (email: cfiggatt@umd.edu).


\textbf{Declaration of competing financial interests} C.M. is co-founder and Chief Scientist at IonQ, Inc.

\end{acknowledgments}

\section{\normalsize Methods}

\section{Constraint Problem for Optimal Parallel Operations} 

Here, we discuss in detail the constraint problem constructed when performing operations in parallel between two ion pairs ($M=2$), as is the case for the data shown in this paper. In order to perform parallel entangling operations involving 2 pairs of qubits $(i,j)$ and $(m,n)$ in a chain of $N$ ions with $N$ motional modes $\omega_k$, the evolution operator \cite{ZhuGates06,Zhu06PRL,Choi14} here can be written as:
\begin{align} \label{eqn:ParallelUnitaryExpanded} 
U_{||}(\tau) &= \exp\left( \sum_{i=0}^{4} \phi_i(\tau) \sigma_i^x + i \sum_{i<j}^{4} \chi_{ij}(\tau) \sigma_i^x \sigma_j^x \right) \nonumber\\
	&= \exp\left( \phi_i(\tau) \sigma_i^x + \phi_j(\tau) \sigma_j^x + \phi_m(\tau) \sigma_m^x  \right. \nonumber\\
	&\qquad  \left. {} + \phi_n(\tau) \sigma_n^x +  i\left[ \chi_{ij}(\tau) \sigma_i^x \sigma_j^x + \chi_{mn}(\tau) \sigma_m^x \sigma_n^x \right. \right. \nonumber\\
	&\qquad \left. \left. {} + \chi_{im}(\tau) \sigma_i^x \sigma_m^x + \chi_{in}(\tau) \sigma_i^x \sigma_n^x \right. \right. \nonumber\\
	&\qquad \left. \left. {} + \chi_{jm}(\tau) \sigma_j^x \sigma_m^x + \chi_{jn}(\tau) \sigma_j^x \sigma_n^x \right] \right) .
\end{align}

At the end of the gate, the $4N$ spin-motion terms (4 ions, $N$ modes) $\hat{\phi}(\tau) = \sum_k ( \alpha_{i,k}(\tau) \hat{a}_k^{\dag} - \alpha_{i,k}^*(\tau) \hat{a}_k )$ 
must go to zero, ensuring all mode trajectories close. 
We wish to entangle only the 2 ion pairs $(i,j)$ and $(m,n)$; since we do not wish to entangle the 4 crosstalk pairs $(i,m)$, $(i,n)$, and so on, they are set to 0. 
The entangling pairs require $\chi = \chi^{\text{ideal}}$, where $ \chi^{\text{ideal}}=\frac{\pi}{4}$ implements a maximally entangling $\xxgate$ gate, and smaller positive values implement partially-entangling gates. 
This then yields a set of $4N+6$ parameters to control for when optimizing pulse sequences to implement parallel $\xxgate$ gates:
\begin{align} \label{eqn:ParallelConstraints}
\alpha_{\{i,j,m,n\},k }(\tau) &= 0 \nonumber\\
\chi_{ij}(\tau) &= \chi_{ij}^{\text{ideal}} \nonumber\\
\chi_{mn}(\tau) &= \chi_{mn}^{\text{ideal}} \nonumber\\
\chi_{im}(\tau) = \chi_{in}(\tau) = \chi_{jm}(\tau) = \chi_{jn}(\tau) &= 0. 
\end{align}

To provide optimal control during the gate and fulfill the constraints in Equation \ref{eqn:ParallelConstraints}, we divide up the gate amplitude $\Omega_i(t)$ into $S$ segments of equal duration $\tau/S$, and vary the amplitude in each segment $\Omega_s$ as an independent variable. In order to implement independent $\xxgate$ gates, we perform distinct signals on the two ion pairs we want to entangle; ions $(i,j)$ see one pulse shape, while ions $(m,n)$ see another. 
The gate amplitude on a given ion then becomes a piecewise-constant function,
\begin{align} \label{eqn:OmegaPiecewise}
\Omega_i(t) = 
  \begin{cases} 
   \Omega_1 & 0 \leq t < \tau/S \\
   \Omega_2 & \tau/S \leq t < 2\tau/S \\
	\vdots & \vdots\\
   \Omega_s & (s-1)\tau/S \leq t < s\tau/S\\
	\vdots & \vdots\\
   \Omega_S & (S-1)\tau/S \leq t < \tau.
  \end{cases}
\end{align}
The accumulated displacement of the spin-motion parameter between ion $i$ and mode $k$, $\alpha_{i,k}(\tau, \Omega_i(t))$, is  
\begin{equation}\label{eqn:alpha1}
\alpha_{i,k}(\tau) = \int_0^{\tau} \eta_{i,k} \Omega_i(t) \sin(\mu t) e^{i\omega_k t} dt,
\end{equation}
where $\eta_{i,k}$ is the spin-motion coupling or Lamb-Dicke parameter. 
The spin-spin interaction term $\chi_{ij}(\tau,\Omega_i(t),\Omega_j(t))$ is 
\begin{align}\label{eqn:chi1}
\chi_{ij}(\tau) &= 2 \int_0^{\tau} dt' \int_0^{t'} dt \sum_k \eta_{i,k} \eta_{j,k} \Omega_i(t) \Omega_j(t)  \nonumber\\
& \qquad {} \times \sin(\mu t) \sin(\mu t') \sin(\omega_k (t'-t)) .
\end{align}

Applying Equation \ref{eqn:OmegaPiecewise}, Equations \ref{eqn:alpha1} and \ref{eqn:chi1} can be re-written as
\begin{align}
\alpha_{i,k}(\tau) &= \sum_{s=1}^S \Omega_s C^i_{k,s}
\end{align}
and
\begin{align}
\chi_{ij}(\tau) &= \sum_{s=1}^S \sum_{s'=1}^S \Omega_s \Omega_{s'} D_{s,s'},
\end{align}
where the terms 
\begin{align}
C^i_{k,s} &= \eta_{i,k} \int_{(s-1)\tau/S}^{s\tau/S} \sin(\mu t) e^{i\omega_k t} dt 
\end{align}
and 
\begin{align} \label{eqn:Dintegral}
D^{ij}_{s,s'} &= \int_{(s-1)\tau/S}^{s\tau/S} dt' \int_{(s'-1)\tau/S}^{s'\tau/S} dt \sum_k \eta_{i,k} \eta_{j,k} \nonumber\\
&\qquad  {} \times \sin(\mu t) \sin(\mu t') \sin(\omega_k (t'-t))
\end{align}
are pre-calculated constants that are functions only of the motional mode frequencies $\omega_k$, the detuning $\mu$, and the segment number $s$, and so can be arranged into $S \times N$ and $S \times S$ matrices for each ion or ion pair, respectively. 
Note that the time ordering of the double integral in Equation \ref{eqn:Dintegral} requires that $t<t'$, so the time-segmented scheme requires $s \leq s'$. In the case $s=s'$, we must force the $t<t'$ constraint, yielding
\begin{align}
D^{ij}_{s=s'} &= \int_{(s-1)\tau/S}^{s\tau/S} dt' \int_{(s'-1)\tau/S}^{t'} dt \sum_k \eta_{i,k} \eta_{j,k} \nonumber\\
&\qquad  {} \times \sin(\mu t) \sin(\mu t') \sin(\omega_k (t'-t)).
\end{align}

If we now arrange the segment amplitudes $\Omega_s$ into two vectors $\mathbf{\Omega_{ij}}$ and $\mathbf{\Omega_{mn}}$, one for each entangling pair $(i,j)$ and $(m,n)$, we can now write our constraint equations from Equation \ref{eqn:ParallelConstraints} as
\begin{align} \label{eqn:ParallelConstraintsVectors}
\begin{pmatrix} \mathbf{C^i}\\ \mathbf{C^j}\\ \mathbf{C^m}\\ \mathbf{C^n}\\ \end{pmatrix} \begin{pmatrix} \mathbf{\Omega_{ij}}\\ \mathbf{\Omega_{mn}} \end{pmatrix} &= 0 \nonumber\\
\begin{pmatrix} \mathbf{\Omega_{ij}^T} & \mathbf{\Omega_{mn}^T} \end{pmatrix} \begin{pmatrix} \mathbf{D^{ij}} & 0\\ 0 & 0 \end{pmatrix} \begin{pmatrix} \mathbf{\Omega_{ij}}\\ \mathbf{\Omega_{mn}} \end{pmatrix} &= \chi_{ij}^{\text{ideal}} \nonumber\\
\begin{pmatrix} \mathbf{\Omega_{ij}^T} & \mathbf{\Omega_{mn}^T} \end{pmatrix} \begin{pmatrix} 0 & 0\\ 0 & \mathbf{D^{mn}} \end{pmatrix} \begin{pmatrix} \mathbf{\Omega_{ij}}\\ \mathbf{\Omega_{mn}} \end{pmatrix} &= \chi_{mn}^{\text{ideal}} \nonumber\\
\begin{pmatrix} \mathbf{\Omega_{ij}^T} & \mathbf{\Omega_{mn}^T} \end{pmatrix} \begin{pmatrix} 0 & \mathbf{D^{\textbf{cross}}}\\ 0 & 0 \end{pmatrix} \begin{pmatrix} \mathbf{\Omega_{ij}}\\ \mathbf{\Omega_{mn}} \end{pmatrix} &= 0, 
\end{align}
where $\left\{ \mathbf{C^i},\mathbf{C^j},\mathbf{C^m},\mathbf{C^n} \right\}$ are the $S \times N$ spin-motion interaction matricies for each segment on each ion, $\left\{ \mathbf{D^{ij}}, \mathbf{D^{mn}} \right\}$ are the two $S \times S$ spin-spin interaction matricies for each segment on the entangling ion pairs, and $\mathbf{D^{\textbf{cross}}} = \left\{\mathbf{D^{im}},\mathbf{D^{in}},\mathbf{D^{jm}},\mathbf{D^{jn}} \right\}$ are the four $S \times S$ spin-spin interaction matricies for each segment on the crosstalk ion pairs. Here we see that while the $C$-constraints are linear, the 6 $D$-constraints on the spin-spin interaction terms are not. 

\section{Optimization Methods}

The constraint problem on $\alpha$ and $\chi$ was converted to an unconstrained optimization problem using the penalty method, specifically by minimizing an objective function that is quadratic in the deviations of $\alpha$ and $\chi$ from their ideal values. This objective function also penalized high power pulse sequences. The objective function was minimized using the built-in MATLAB function “fminunc”.  The initial guess used for the pulse shapes in the optimization protocol was that one pair would have an all-positive-amplitude shape, while the other pair would see positive amplitudes for the first half, and negative amplitudes for the second half of the gate.

\section{Fidelity of Parallel $\xxgate$ Operations}

Here, we calculate the fidelity of simultaneous $\xxgate$ gate operations as a function of the above control parameters. The fidelity is given by
\begin{equation} \label{eqn:ParallelFidelitySetup}
F_{||} = \bra{\psi_{\text{init}}} U^{\dag}_{\text{ideal}} \rho_r U_{\text{ideal}} \ket{\psi_{\text{init}}},
\end{equation}
where $\rho_r$ is the density matrix for the experimental operation traced over the motion,
\begin{equation} \label{eqn:RhoR}
\rho_r =  Tr_m\left[ U_{\text{expt}} \ket{\psi_{\text{init}}} \bra{\psi_{\text{init}}} U^{\dag}_{\text{expt}} \right].
\end{equation} 
Plugging in all values and solving, we derive the fidelity,
\onecolumngrid
\allowdisplaybreaks
\begin{align} \label{eqn:ParallelFidelity}
& F_{||}\left(\alpha_{\{i,j,m,n\},k }, \chi_{ij}, \chi_{ij}^{\text{ideal}}, \chi_{mn}, \chi_{mn}^{\text{ideal}}, \chi_{im}, \chi_{in}, \chi_{jm}, \chi_{jn} \right) = \nonumber\\
& \frac{1}{128} \left( 8 + \Gamma_{+---} + \Gamma_{+--+} + \Gamma_{+-+-} + \Gamma_{+-++} + \Gamma_{++--} + \Gamma_{++-+} + \Gamma_{+++-} + \Gamma_{++++}  \right.\nonumber \\
&\qquad \left. {} + 2\left(\Gamma_{0+--} + \Gamma_{+000}\right)\cos\left[2\left(\Delta\chi_{ij} - \chi_{im} - \chi_{in}\right)\right] + 2\left(\Gamma_{0++-} + \Gamma_{+000}\right)\cos\left[2\left(\Delta\chi_{ij} + \chi_{im} - \chi_{in}\right)\right]  \right.\nonumber \\
&\qquad \left. {} + 2\left(\Gamma_{0+-+} + \Gamma_{+000}\right)\cos\left[2\left(\Delta\chi_{ij} - \chi_{im} + \chi_{in}\right)\right] + 2\left(\Gamma_{0+++} + \Gamma_{+000}\right)\cos\left[2\left(\Delta\chi_{ij} + \chi_{im} + \chi_{in}\right)\right]  \right.\nonumber \\
&\qquad \left. {} + 2\left(\Gamma_{0+00} + \Gamma_{+0--}\right)\cos\left[2\left(\Delta\chi_{ij} - \chi_{jm} - \chi_{jn}\right)\right] + 2\left(\Gamma_{0+00} + \Gamma_{+0+-}\right)\cos\left[2\left(\Delta\chi_{ij} + \chi_{jm} - \chi_{jn}\right)\right] \right.\nonumber \\
&\qquad \left. {} + 2\left(\Gamma_{0+00} + \Gamma_{+0-+}\right)\cos\left[2\left(\Delta\chi_{ij} - \chi_{jm} + \chi_{jn}\right)\right]  + 2\left(\Gamma_{0+00} + \Gamma_{+0++}\right)\cos\left[2\left(\Delta\chi_{ij} + \chi_{jm} + \chi_{jn}\right)\right] \right.\nonumber \\
&\qquad \left. {} + 2\left(\Gamma_{00+0} + \Gamma_{+-0+}\right)\cos\left[2\left(\chi_{im} - \chi_{jm} + \Delta\chi_{mn}\right)\right] + 2\left(\Gamma_{00+0} + \Gamma_{++0+}\right)\cos\left[2\left(\chi_{im} + \chi_{jm} + \Delta\chi_{mn}\right)\right] \right.\nonumber \\
&\qquad \left. {} + 2\left(\Gamma_{00+0} + \Gamma_{+-0-}\right)\cos\left[2\left(\chi_{im} - \chi_{jm} - \Delta\chi_{mn}\right)\right] + 2\left(\Gamma_{00+0} + \Gamma_{++0-}\right)\cos\left[2\left(\chi_{im} + \chi_{jm} - \Delta\chi_{mn}\right)\right] \right.\nonumber \\
&\qquad \left. {} + 2\left(\Gamma_{000+} + \Gamma_{+-+0}\right)\cos\left[2\left(\chi_{in} - \chi_{jn} + \Delta\chi_{mn}\right)\right] + 2\left(\Gamma_{000+} + \Gamma_{+++0}\right)\cos\left[2\left(\chi_{in} + \chi_{jn} + \Delta\chi_{mn}\right)\right] \right.\nonumber \\
&\qquad \left. {} + 2\left(\Gamma_{000+} + \Gamma_{+--0}\right)\cos\left[2\left(\chi_{in} - \chi_{jn} - \Delta\chi_{mn}\right)\right] + 2\left(\Gamma_{000+} + \Gamma_{++-0}\right)\cos\left[2\left(\chi_{in} + \chi_{jn} - \Delta\chi_{mn}\right)\right] \right.\nonumber \\
&\qquad \left. {} + 2\left(\Gamma_{00++} + \Gamma_{+-00}\right)\cos\left[2\left(\chi_{im} + \chi_{in} - \chi_{jm} - \chi_{jn}\right)\right]  \right.\nonumber \\
&\qquad \left. {} + 2\left(\Gamma_{00+-} + \Gamma_{++00}\right)\cos\left[2\left(\chi_{im} - \chi_{in} + \chi_{jm} - \chi_{jn}\right)\right]  \right.\nonumber \\
&\qquad \left. {} + 2\left(\Gamma_{00+-} + \Gamma_{+-00}\right)\cos\left[2\left(\chi_{im} - \chi_{in} - \chi_{jm} + \chi_{jn}\right)\right]  \right.\nonumber \\
&\qquad \left. {} + 2\left(\Gamma_{00++} + \Gamma_{++00}\right)\cos\left[2\left(\chi_{im} + \chi_{in} + \chi_{jm} + \chi_{jn}\right)\right]  \right.\nonumber \\
&\qquad \left. {} + 2\left(\Gamma_{0+0-} + \Gamma_{+0-0}\right)\cos\left[2\left(\Delta\chi_{ij} - \chi_{in} - \chi_{jm} + \Delta\chi_{mn}\right)\right]  \right.\nonumber \\
&\qquad \left. {} + 2\left(\Gamma_{0+0+} + \Gamma_{+0+0}\right)\cos\left[2\left(\Delta\chi_{ij} + \chi_{in} + \chi_{jm} + \Delta\chi_{mn}\right)\right]  \right.\nonumber \\
&\qquad \left. {} + 2\left(\Gamma_{0+0+} + \Gamma_{+0-0}\right)\cos\left[2\left(\Delta\chi_{ij} + \chi_{in} - \chi_{jm} - \Delta\chi_{mn}\right)\right]  \right.\nonumber \\
&\qquad \left. {} + 2\left(\Gamma_{0+0-} + \Gamma_{+0+0}\right)\cos\left[2\left(\Delta\chi_{ij} - \chi_{in} + \chi_{jm} - \Delta\chi_{mn}\right)\right]  \right.\nonumber \\
&\qquad \left. {} + 2\left(\Gamma_{0+-0} + \Gamma_{+00-}\right)\cos\left[2\left(\Delta\chi_{ij} - \chi_{im} - \chi_{jn} + \Delta\chi_{mn}\right)\right]  \right.\nonumber \\
&\qquad \left. {} + 2\left(\Gamma_{0++0} + \Gamma_{+00+}\right)\cos\left[2\left(\Delta\chi_{ij} + \chi_{im} + \chi_{jn} + \Delta\chi_{mn}\right)\right]  \right.\nonumber \\
&\qquad \left. {} + 2\left(\Gamma_{0++0} + \Gamma_{+00-}\right)\cos\left[2\left(\Delta\chi_{ij} + \chi_{im} - \chi_{jn} - \Delta\chi_{mn}\right)\right]  \right.\nonumber \\
&\qquad \left. {} + 2\left(\Gamma_{0+-0} + \Gamma_{+00+}\right)\cos\left[2\left(\Delta\chi_{ij} - \chi_{im} + \chi_{jn} - \Delta\chi_{mn}\right)\right]   
\right).
\end{align} 
\allowdisplaybreaks[0]
Here,
\begin{equation}
\Gamma_{A_i A_j A_m A_n} = \exp \left(-\frac{1}{2} \sum_k \beta_k \left| 2\left( A_i \alpha_{i,k} + A_j \alpha_{j,k} + A_m \alpha_{m,k} + A_n \alpha_{n,k} \right) \right|^2 \right),
\end{equation}
\twocolumngrid

where the parameters $\{ A_i, A_j, A_m, A_n \}$ can be $\{0, \pm 1\}$, and we indicate them as $\{ 0 \rightarrow 0, +1 \rightarrow +, -1 \rightarrow - \}$ in Equations \ref{eqn:ParallelFidelity} and \ref{eqn:GHZFidelity} for display purposes. The inverse mode temperature $\beta_k$ is
\begin{equation}
\beta_k = \coth \left( \frac{\hbar \omega_k}{k_B T} \right) = \coth \left[\frac{1}{2} \ln \left( 1+ \frac{1}{\bar{n}_k} \right) \right],
\end{equation}
where $\bar{n}_k$ is the average phonon number in the $k^{\text{th}}$ mode. We additionally use
\begin{align}
\Delta\chi_{ij} &= \chi_{ij} - \chi_{ij}^{\text{ideal}} \nonumber\\
\Delta\chi_{mn} &= \chi_{mn} - \chi_{mn}^{\text{ideal}}.
\end{align}
Plugging in the ideal-case parameters, where $\alpha_{\{i,j,m,n\},k }(\tau)=0$, $\chi_{im}=\chi_{in}=\chi_{jm}=\chi_{jn}=0$, $\chi_{ij} = \chi_{ij}^{\text{ideal}}$, and $\chi_{mn} = \chi_{mn}^{\text{ideal}}$, we indeed get $F_{||}=1$. See \cite{FiggattThesis18} for a more detailed derivation.

\vspace{5 mm}

\section{Toward a Single-Operation GHZ State}

This control scheme for parallel 2-qubit entangling gates in ions also suggests a method for performing multi-qubit entanglement in a single operation. Of particular interest is the creation of GHZ states \cite{GHZ89}, which are a class of non-biseparable maximally-entangled multi-qubit states.  
Setting all 6 spin-spin interaction terms to $\chi = \frac{\pi}{4}$ in Equation \ref{eqn:ParallelUnitary} yields the unitary
\onecolumngrid
\setcounter{MaxMatrixCols}{20}
\begin{align}
U_{\text{GHZ}}^{\text{ideal}} &= U \left( \alpha_{\{i,j,m,n\},k }=0,  \chi_{ij}= \chi_{im}= \chi_{in}= \chi_{jm}= \chi_{jn}= \chi_{mn} = \frac{\pi}{4}\right) \nonumber\\
&=  \frac{1}{\sqrt{2}} \left( \mathbb{I} - i \left(\sigma^x \right)^{\otimes 4} \right) \nonumber\\
&= \frac{1}{\sqrt{2}}\begin{pmatrix}
1  & 0  & 0  & 0  & 0  & 0  & 0  & 0  & 0  & 0  & 0  & 0  & 0  & 0  & 0  & -i \\
0  & 1  & 0  & 0  & 0  & 0  & 0  & 0  & 0  & 0  & 0  & 0  & 0  & 0  & -i & 0  \\
0  & 0  & 1  & 0  & 0  & 0  & 0  & 0  & 0  & 0  & 0  & 0  & 0  & -i & 0  & 0  \\
0  & 0  & 0  & 1  & 0  & 0  & 0  & 0  & 0  & 0  & 0  & 0  & -i & 0  & 0  & 0  \\
0  & 0  & 0  & 0  & 1  & 0  & 0  & 0  & 0  & 0  & 0  & -i & 0  & 0  & 0  & 0  \\
0  & 0  & 0  & 0  & 0  & 1  & 0  & 0  & 0  & 0  & -i & 0  & 0  & 0  & 0  & 0  \\
0  & 0  & 0  & 0  & 0  & 0  & 1  & 0  & 0  & -i & 0  & 0  & 0  & 0  & 0  & 0  \\
0  & 0  & 0  & 0  & 0  & 0  & 0  & 1  & -i & 0  & 0  & 0  & 0  & 0  & 0  & 0  \\
0  & 0  & 0  & 0  & 0  & 0  & 0  & -i & 1  & 0  & 0  & 0  & 0  & 0  & 0  & 0  \\
0  & 0  & 0  & 0  & 0  & 0  & -i & 0  & 0  & 1  & 0  & 0  & 0  & 0  & 0  & 0  \\
0  & 0  & 0  & 0  & 0  & -i & 0  & 0  & 0  & 0  & 1  & 0  & 0  & 0  & 0  & 0  \\
0  & 0  & 0  & 0  & -i & 0  & 0  & 0  & 0  & 0  & 0  & 1  & 0  & 0  & 0  & 0  \\
0  & 0  & 0  & -i & 0  & 0  & 0  & 0  & 0  & 0  & 0  & 0  & 1  & 0  & 0  & 0  \\
0  & 0  & -i & 0  & 0  & 0  & 0  & 0  & 0  & 0  & 0  & 0  & 0  & 1  & 0  & 0  \\
0  & -i & 0  & 0  & 0  & 0  & 0  & 0  & 0  & 0  & 0  & 0  & 0  & 0  & 1  & 0  \\
-i & 0  & 0  & 0  & 0  & 0  & 0  & 0  & 0  & 0  & 0  & 0  & 0  & 0  & 0  & 1 
\end{pmatrix},
\end{align}
and adding one $Z$ rotation $R_z\left( \theta \right) = \begin{pmatrix} e^{-i\frac{\theta}{2}} & 0 \\ 0 & e^{i\frac{\theta}{2}} \end{pmatrix}$ with $\theta = \frac{\pi}{2}$ produces a 4-qubit GHZ state:
\begin{align}
R_z^1\left(\frac{\pi}{2} \right) \cdot U_{\text{GHZ}}^{\text{ideal}} \ket{0000} &= \frac{1}{\sqrt{2}} \left(\ket{0000} + \ket{1111} \right).
\end{align}

Following a similar derivation as with parallel gates, we therefore calculate the objective fidelity function to be 
\allowdisplaybreaks
\begin{align} \label{eqn:GHZFidelity}
& F_{\text{GHZ}}\left(\alpha_{\{i,j,m,n\},k }, \chi_{ij}^{\text{ideal}}, \chi_{im}^{\text{ideal}}, \chi_{in}^{\text{ideal}}, \chi_{jm}^{\text{ideal}}, \chi_{jn}^{\text{ideal}}, \chi_{mn}^{\text{ideal}}, \chi_{ij}, \chi_{im}, \chi_{in}, \chi_{jm}, \chi_{jn}, \chi_{mn} \right) = \nonumber\\
& \frac{1}{128} \left( 8 + \Gamma_{+---} + \Gamma_{+--+} + \Gamma_{+-+-} + \Gamma_{+-++} + \Gamma_{++--} + \Gamma_{++-+} + \Gamma_{+++-} + \Gamma_{++++}  \right.\nonumber \\
&\qquad \left. {} + 2\left(\Gamma_{0+++} + \Gamma_{+000}\right)\cos\left[2\left(\Delta\chi_{ij} + \Delta\chi_{im} + \Delta\chi_{in}\right)\right] + 2\left(\Gamma_{0+-+} + \Gamma_{+000}\right)\cos\left[2\left(\Delta\chi_{ij} - \Delta\chi_{im} + \Delta\chi_{in}\right)\right]  \right.\nonumber \\
&\qquad \left. {} + 2\left(\Gamma_{0++-} + \Gamma_{+000}\right)\cos\left[2\left(\Delta\chi_{ij} + \Delta\chi_{im} - \Delta\chi_{in}\right)\right]  + 2\left(\Gamma_{0+--} + \Gamma_{+000}\right)\cos\left[2\left(\Delta\chi_{ij} - \Delta\chi_{im} - \Delta\chi_{in}\right)\right]  \right.\nonumber \\
&\qquad \left. {} + 2\left(\Gamma_{0+00} + \Gamma_{+0++}\right)\cos\left[2\left(\Delta\chi_{ij} + \Delta\chi_{jm} + \Delta\chi_{jn}\right)\right] + 2\left(\Gamma_{0+00} + \Gamma_{+0-+}\right)\cos\left[2\left(\Delta\chi_{ij} - \Delta\chi_{jm} + \Delta\chi_{jn}\right)\right] \right.\nonumber \\
&\qquad \left. {} + 2\left(\Gamma_{0+00} + \Gamma_{+0+-}\right)\cos\left[2\left(\Delta\chi_{ij} + \Delta\chi_{jm} - \Delta\chi_{jn}\right)\right] + 2\left(\Gamma_{0+00} + \Gamma_{+0--}\right)\cos\left[2\left(\Delta\chi_{ij} - \Delta\chi_{jm} - \Delta\chi_{jn}\right)\right]  \right.\nonumber \\
&\qquad \left. {} + 2\left(\Gamma_{00+0} + \Gamma_{++0+}\right)\cos\left[2\left(\Delta\chi_{im} + \Delta\chi_{jm} + \Delta\chi_{mn}\right)\right] + 2\left(\Gamma_{00+0} + \Gamma_{+-0+}\right)\cos\left[2\left(\Delta\chi_{im} - \Delta\chi_{jm} + \Delta\chi_{mn}\right)\right] \right.\nonumber \\
&\qquad \left. {} + 2\left(\Gamma_{00+0} + \Gamma_{++0-}\right)\cos\left[2\left(\Delta\chi_{im} + \Delta\chi_{jm} - \Delta\chi_{mn}\right)\right] + 2\left(\Gamma_{00+0} + \Gamma_{+-0-}\right)\cos\left[2\left(\Delta\chi_{im} - \Delta\chi_{jm} - \Delta\chi_{mn}\right)\right] \right.\nonumber \\
&\qquad \left. {} + 2\left(\Gamma_{000+} + \Gamma_{+++0}\right)\cos\left[2\left(\Delta\chi_{in} + \Delta\chi_{jn} + \Delta\chi_{mn}\right)\right] + 2\left(\Gamma_{000+} + \Gamma_{+-+0}\right)\cos\left[2\left(\Delta\chi_{in} - \Delta\chi_{jn} + \Delta\chi_{mn}\right)\right] \right.\nonumber \\
&\qquad \left. {} + 2\left(\Gamma_{000+} + \Gamma_{++-0}\right)\cos\left[2\left(\Delta\chi_{in} + \Delta\chi_{jn} - \Delta\chi_{mn}\right)\right] + 2\left(\Gamma_{000+} + 2\Gamma_{+--0}\right)\cos\left[2\left(\Delta\chi_{in} - \Delta\chi_{jn} - \Delta\chi_{mn}\right)\right] \right.\nonumber \\
&\qquad \left. {} + 2\left(\Gamma_{00++} + \Gamma_{++00}\right)\cos\left[2\left(\Delta\chi_{im} + \Delta\chi_{in} + \Delta\chi_{jm} + \Delta\chi_{jn}\right)\right]  \right.\nonumber \\
&\qquad \left. {} + 2\left(\Gamma_{00+-} + \Gamma_{+-00}\right)\cos\left[2\left(\Delta\chi_{im} - \Delta\chi_{in} - \Delta\chi_{jm} + \Delta\chi_{jn}\right)\right]  \right.\nonumber \\
&\qquad \left. {} + 2\left(\Gamma_{00+-} + \Gamma_{++00}\right)\cos\left[2\left(\Delta\chi_{im} - \Delta\chi_{in} + \Delta\chi_{jm} - \Delta\chi_{jn}\right)\right]  \right.\nonumber \\
&\qquad \left. {} + 2\left(\Gamma_{00++} + \Gamma_{+-00}\right)\cos\left[2\left(\Delta\chi_{im} + \Delta\chi_{in} - \Delta\chi_{jm} - \Delta\chi_{jn}\right)\right]  \right.\nonumber \\
&\qquad \left. {} + 2\left(\Gamma_{0+0+} + \Gamma_{+0+0}\right)\cos\left[2\left(\Delta\chi_{ij} + \Delta\chi_{in} + \Delta\chi_{jm} + \Delta\chi_{mn}\right)\right]  \right.\nonumber \\
&\qquad \left. {} + 2\left(\Gamma_{0+0-} + \Gamma_{+0-0}\right)\cos\left[2\left(\Delta\chi_{ij} - \Delta\chi_{in} - \Delta\chi_{jm} + \Delta\chi_{mn}\right)\right]  \right.\nonumber \\
&\qquad \left. {} + 2\left(\Gamma_{0+0-} + \Gamma_{+0+0}\right)\cos\left[2\left(\Delta\chi_{ij} - \Delta\chi_{in} + \Delta\chi_{jm} - \Delta\chi_{mn}\right)\right]  \right.\nonumber \\
&\qquad \left. {} + 2\left(\Gamma_{0+0+} + \Gamma_{+0-0}\right)\cos\left[2\left(\Delta\chi_{ij} + \Delta\chi_{in} - \Delta\chi_{jm} - \Delta\chi_{mn}\right)\right]  \right.\nonumber \\
&\qquad \left. {} + 2\left(\Gamma_{0++0} + \Gamma_{+00+}\right)\cos\left[2\left(\Delta\chi_{ij} + \Delta\chi_{im} + \Delta\chi_{jn} + \Delta\chi_{mn}\right)\right]  \right.\nonumber \\
&\qquad \left. {} + 2\left(\Gamma_{0+-0} + \Gamma_{+00-}\right)\cos\left[2\left(\Delta\chi_{ij} - \Delta\chi_{im} - \Delta\chi_{jn} + \Delta\chi_{mn}\right)\right]  \right.\nonumber \\
&\qquad \left. {} + 2\left(\Gamma_{0+-0} + \Gamma_{+00+}\right)\cos\left[2\left(\Delta\chi_{ij} - \Delta\chi_{im} + \Delta\chi_{jn} - \Delta\chi_{mn}\right)\right]  \right.\nonumber \\
&\qquad \left. {} + 2\left(\Gamma_{0++0} + \Gamma_{+00-}\right)\cos\left[2\left(\Delta\chi_{ij} + \Delta\chi_{im} - \Delta\chi_{jn} - \Delta\chi_{mn}\right)\right]  
\right),
\end{align}
\allowdisplaybreaks[0]
\twocolumngrid
where 
\begin{align}
\Delta\chi_{ij} &= \chi_{ij} - \chi_{ij}^{\text{ideal}} \nonumber\\
\Delta\chi_{mn} &= \chi_{mn} - \chi_{mn}^{\text{ideal}} \nonumber\\
\Delta\chi_{im} &= \chi_{im} - \chi_{im}^{\text{ideal}} \nonumber\\
\Delta\chi_{in} &= \chi_{in} - \chi_{in}^{\text{ideal}} \nonumber\\
\Delta\chi_{jm} &= \chi_{jm} - \chi_{jm}^{\text{ideal}} \nonumber\\
\Delta\chi_{jn} &= \chi_{jn} - \chi_{jn}^{\text{ideal}}.
\end{align}

This indicates we may be able to use the same optimization approach to produce pulse shapes that will create GHZ states when applied to the ions. Unlike with parallel gates, however, it may be necessary to allow independent pulse shapes on all 4 ions, rather than solving for pairwise solutions; this will provide more free parameters. Additional challenges will include finding effective calibration techniques when implementing such gates on the experiment, since there will be 6 interactions that will all need to be at the same strength, but only 4 control signals. Our current approach of calibrating a 2-qubit gate by adjusting the overall power for the pulse shape applied by the control signal may no longer work; new techniques with more degrees of freedom may be needed, such as independently adjusting the power for a few different sections of the pulse shape on each ion.

The benefits of implementing GHZ states with fewer gates would be significant, as it would substantially reduce the circuit depth of several important algorithms. While the use of axial modes for multi-qubit GHZ states has already been shown \cite{MonzGHZ11}, this scheme represents a new method for use with radial mode interactions. With only 2-qubit gates available, building a GHZ state of size $N$ requires $O(N)$ 2-qubit gates. With parallel 2-qubit gates available, the gate depth required to build a GHZ state is reduced to $O(\log(N))$; this is accomplished with a binary tree algorithm by dividing all qubits into pairs and entangling those pairs in parallel, then entangling pairs of these pairs, and so on until all are entangled. A single-operation GHZ state would drop this circuit depth to unity. Single-operation GHZ state construction will greatly enhance the efficiency of several algorithms; for example, arbitrary stabilizer circuits require $O(\frac{N^2}{\log(N)})$ $\cnotgate$ gates \cite{PatelOptimalSynthesis08}, but could be implemented in $O(N)$ gates with single-operation GHZ state circuitry \cite{MaslovGlobalInteractions17}. Single-operation GHZ state creation will also benefit applications such as quantum secret sharing \cite{HillerySecretSharing99}, Toffoli-$N$ gates, the quantum Fourier transform, and quantum Fourier adder circuits \cite{MaslovGlobalInteractions17}.

\section{Experimental Setup}

The experiments are performed on a linear chain of five trapped $^{171}$Yb$^+$ ions that are laser cooled to near their ground state. We designate the qubit as the $|0\rangle \equiv |F=0; m_F=0\rangle$ and $|1\rangle \equiv |F=1; m_F=0\rangle$ hyperfine-split electronic states of the ion's $^2S_{1/2}$ manifold \cite{Olmschenk07}, which are first-order magnetic-field-insensitive clock states with a splitting of 12.642821 GHz. Coherent operations are performed by counterpropagating Raman beams from a single $355\:$nm mode-locked laser. The first Raman beam is a global beam applied to the entire chain, while the second is split into individual addressing beams to target each ion qubit \cite{Debnath16}. Additionally, a multi-channel arbitrary waveform generator (AWG) provides separate RF control signals to each ion's individual addressing beam, providing the individual phase, frequency, and amplitude controls necessary to execute independent 2-qubit operations in parallel. Qubits are initialized to the $\ket{0}$ state using optical pumping, and read out by separate channels of a multi-channel photomultiplier tube (PMT) array using state-dependent fluorescence.

\section{Calculating Fidelities of 2-Qubit Entangling Gates}\label{sec:ParityDerivation}

The fidelity of a 2-qubit $\xxgate(\chi)$ entangling gate can be measured by scanning the phase $\phi$ of a global $\frac{\pi}{2}$ rotation applied after performing the $\xxgate$ gate and calculating the parity at each point of the scan \cite{Sackett4ParticleEntanglement00, KimYbEntanglement09, ManningThesis14}. 
Here, a rotation gate is defined as $R \left( \theta, \phi \right) = \begin{pmatrix} \cos \frac{\theta}{2} & -ie^{-i\phi}\sin\frac{\theta}{2}\\ -ie^{i\phi}\sin\frac{\theta}{2} & \cos \frac{\theta}{2}\end{pmatrix}$.
The analysis pulses used experimentally are rotations using the SK1 composite pulse for increased robustness against errors in the rotation angle \cite{KabytayevSK1_14, BrownSK1_04}. For the four ions involved in each operation, the parity analysis was performed for all 6 possible pairs within the set, allowing for analysis of the 2 entangled ion pairs as well as the 4 crosstalk pairs. 

We start with a global rotation on 2 qubits, 
\begin{align}
R_G \left(\frac{\pi}{2},\phi \right) &= R_1 \left(\frac{\pi}{2},\phi \right) \otimes R_2 \left(\frac{\pi}{2},\phi \right) \nonumber\\[0.3em]
	&= \frac{1}{2}\begin{pmatrix}
1 		 & -ie^{-i\phi} 	& -ie^{-i\phi} 	& -e^{-2i\phi}\\
-ie^{i\phi}  & 1 			& -1 			& -ie^{-i\phi}\\
-ie^{i\phi}  & -1 		& 1 			& -ie^{-i\phi}\\
-e^{2i\phi} & -ie^{i\phi} 	& -ie^{i\phi} 	& 1
\end{pmatrix},
\end{align}
and a general 2-qubit density matrix $\rho_g$ that represents the density matrix produced after experimentally performing an $\xxgate$ gate,
\begin{align}
\rho_g &= \begin{pmatrix}
\rho_{00} 		& \rho_{01} 	& \rho_{02}	& \rho_{03}\\
\rho_{01}^*  	& \rho_{11}	& \rho_{12} 	& \rho_{13}\\
\rho_{02}^*  	& \rho_{12}^*	& \rho_{22}	& \rho_{23}\\
\rho_{03}^* 	& \rho_{13}^* 	& \rho_{23}^* 	& \rho_{33}
\end{pmatrix}, \label{eqn:rhog}
\end{align} 
where $\rho_{00} = \bra{00} \xxgate \ket{00}$, $\rho_{01} = \bra{00} \xxgate \ket{01}$, $\ldots$, $\rho_{23} = \bra{10} \xxgate \ket{11} $, $\rho_{33} = \bra{11} \xxgate \ket{11}$. After performing the analysis pulse $\rgate_G \left(\frac{\pi}{2},\phi \right)$, the new density matrix $\rho_a = \rgate_G \left(\frac{\pi}{2},\phi \right) \cdot \rho_g \cdot R_G^{\dag} \left(\frac{\pi}{2},\phi \right)$ is used to calculate the parity:
\begin{align}
\Pi \left( \rho_a, \phi \right) &= \left( \rho^{00}_a + \rho^{33}_a \right) - \left( \rho^{11}_a + \rho^{22}_a \right) \nonumber\\
	&= 2A_{12} \cos \phi_{12} - 2A_{03} \cos (2\phi - \phi_{03}) \nonumber\\
	&= 2A_{12} \cos \phi_{12} - A_{\Pi} \cos (2\phi - \phi_{03}), \label{eqn:ParityCurve}
\end{align}
where parity is defined as the sum of the even parity populations minus the sum of the odd parity populations and the coherences (off-diagonal density matrix elements) from $\rho_g$ are re-written in the form $\rho_{xy} = A_{xy} e^{-i \phi_{xy}}$. Let us also define the parity amplitude $A_{\Pi} \equiv 2A_{03}$.

Now we calculate the fidelity of an $\xxgate(\chi)$ gate. Using the $\xxgate(\chi)$ gate unitary,
\begin{equation} \label{eqn:XXGateChi}
\xxgate(\chi)=\begin{pmatrix}
\cos(\chi) & 0 & 0 & \hspace{-2mm}-i \sin(\chi)\\[0.3em]
0 & \hspace{-1mm}\cos(\chi) & \hspace{-3mm}-i \sin(\chi) & 0\\[0.3em]
0 & \hspace{-3mm}-i \sin(\chi) & \hspace{-1mm}\cos(\chi) & 0\\[0.3em]
-i \sin(\chi) & 0 & 0 & \cos(\chi)
\end{pmatrix},
\end{equation}
we construct the ideal density matrix after an $\xxgate(\chi)$ gate,
\begin{align}
\rho_{\text{ideal}} &= \xxgate(\chi) \cdot \ket{00} \bra{00} \cdot \xxgate(\chi)^{\dag} \nonumber\\[0.3em]
	&= \begin{pmatrix}
\cos^2(\chi)		& 0 	& 0		 	& i\cos(\chi)\sin(\chi)\\
0 				& 0 	& 0 			& 0\\
0 				& 0 	& 0 			& 0\\
-i\cos(\chi)\sin(\chi)	& 0 	& 0		 	& \sin^2(\chi)
\end{pmatrix}. \label{eqn:rhoideal}
\end{align}
The fidelity of the general fidelity matrix $\rho_g$ with respect to the ideal fidelity matrix $\rho_{\text{ideal}}$ is given by
\begin{equation}
F(\chi) = Tr \left[ \rho_{\text{ideal}}(\chi) \cdot \rho_g \cdot \rho_{\text{ideal}}^{\dag}(\chi) \right].
\end{equation}
Plugging in equations \ref{eqn:rhog} and \ref{eqn:rhoideal}, using $A_{\Pi} \equiv 2A_{03}$, and simplifying yields
\begin{equation} \label{eqn:FXXChi}
F(\chi) = \rho_{00} \cos^2(\chi) + \rho_{33} \sin^2(\chi) + A_{\Pi}\cos(\chi)\sin(\chi)
\end{equation}
as the fidelity of an $\xxgate(\chi)$ gate. Specifically for maximally entangling gates, we use $\chi = \frac{\pi}{4}$ and get
\begin{equation} \label{eqn:FXXpi4}
F \left(\chi=\frac{\pi}{4} \right) = \frac{1}{2} \left( \rho_{00} + \rho_{33} \right) + \frac{1}{2} A_{\Pi}.
\end{equation}
While $\rho_{00}$ and $\rho_{33}$ are simply the populations in $\ket{00}$ and $\ket{11}$ respectively after an $\xxgate$ gate, we still need the $A_{\Pi}$ term. We can extract this from a parity scan using Equation \ref{eqn:ParityCurve}. Given a perfect $\xxgate(\chi)$ gate where $A_{12}=0$, $\phi_{03}= -\frac{\pi}{2}$, and $A_{\Pi} = 2A_{03}=2\cos(\chi)\sin(\chi)$ (from Equation \ref{eqn:rhoideal}), scanning the analysis phase $\phi$ from 0 to $2\pi$ and measuring the parity at each point will yield a sine curve of amplitude $2\cos(\chi)\sin(\chi)$ with 2 periods in the range from 0 to $2\pi$. (For a fully entangling $\xxgate(\chi=\frac{\pi}{4})$ gate, the sine curve should have amplitude 1.) Consequently, by fitting a sine curve to this measured parity curve, we can estimate the parity amplitude $A_{\Pi}$ and use it in Equation \ref{eqn:FXXChi} to calculate the gate fidelity.

\section{Additional Parity Curves and Fidelity Data for 2-Qubit Entangling Gates}\label{sec:MoreParityFlops}

\begin{figure*}
\centering
\begin{tabular}[c]{c c}
\multicolumn{1}{l}{\bf{(a)}} & \multicolumn{1}{l}{\bf{(b)}} \\
\multicolumn{1}{c}{\includegraphics[width=0.5\textwidth,valign=T]{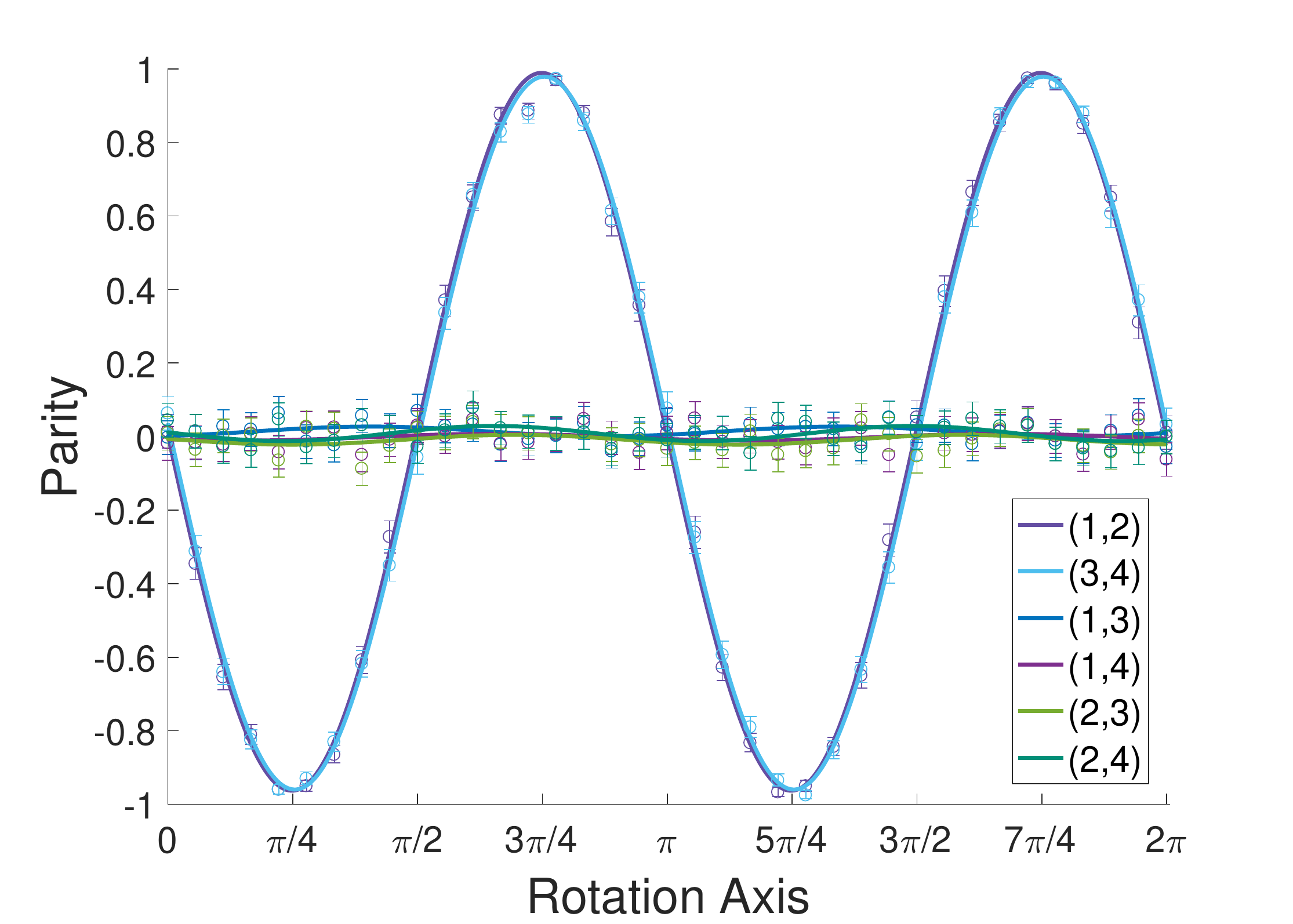}} & \multicolumn{1}{c}{\includegraphics[width=0.5\textwidth,valign=T]{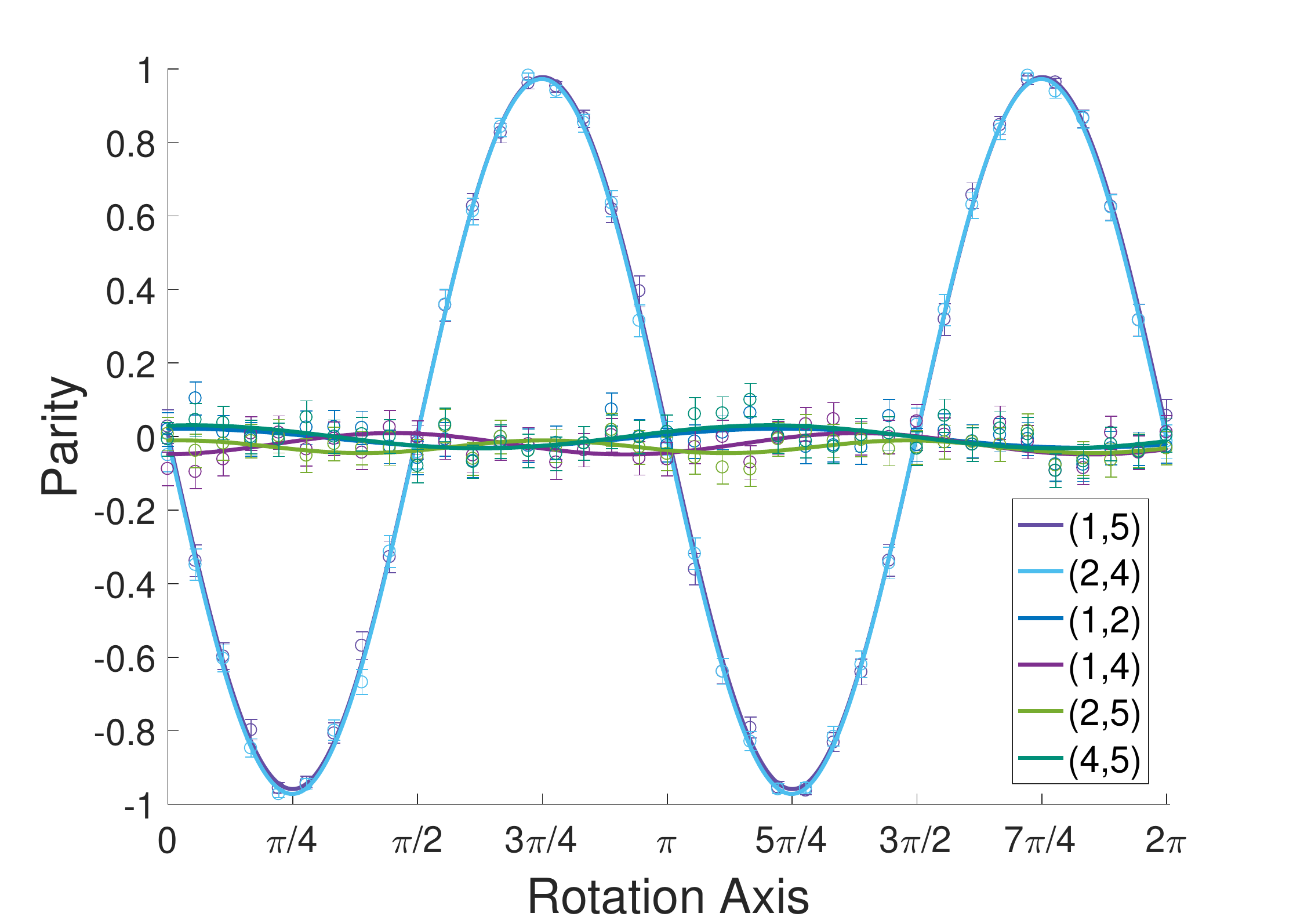}}\\ 
\\
\multicolumn{1}{l}{\bf{(c)}} & \multicolumn{1}{l}{\bf{(d)}} \\
\multicolumn{1}{c}{\includegraphics[width=0.5\textwidth,valign=T]{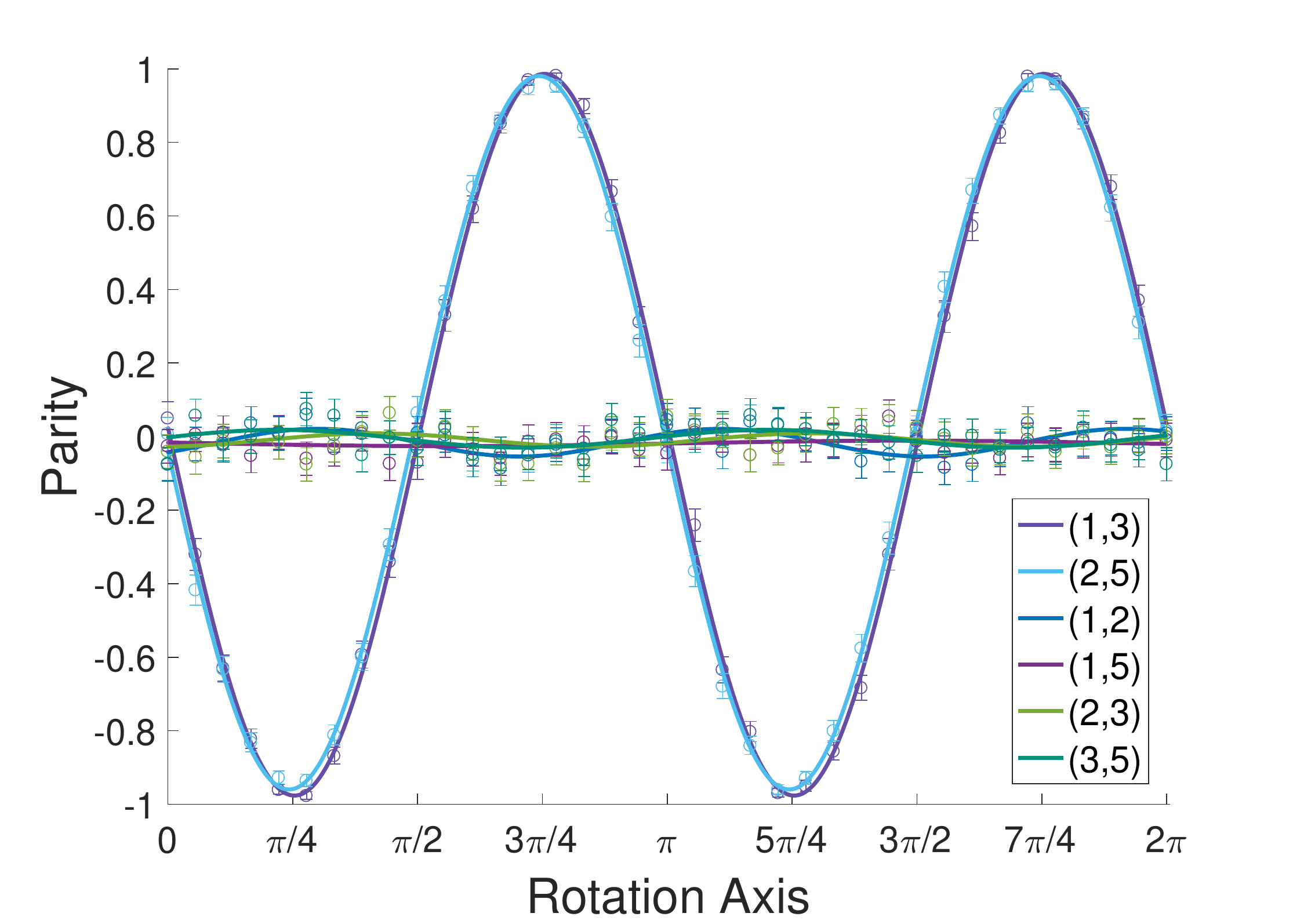}} & \multicolumn{1}{c}{\includegraphics[width=0.5\textwidth,valign=T]{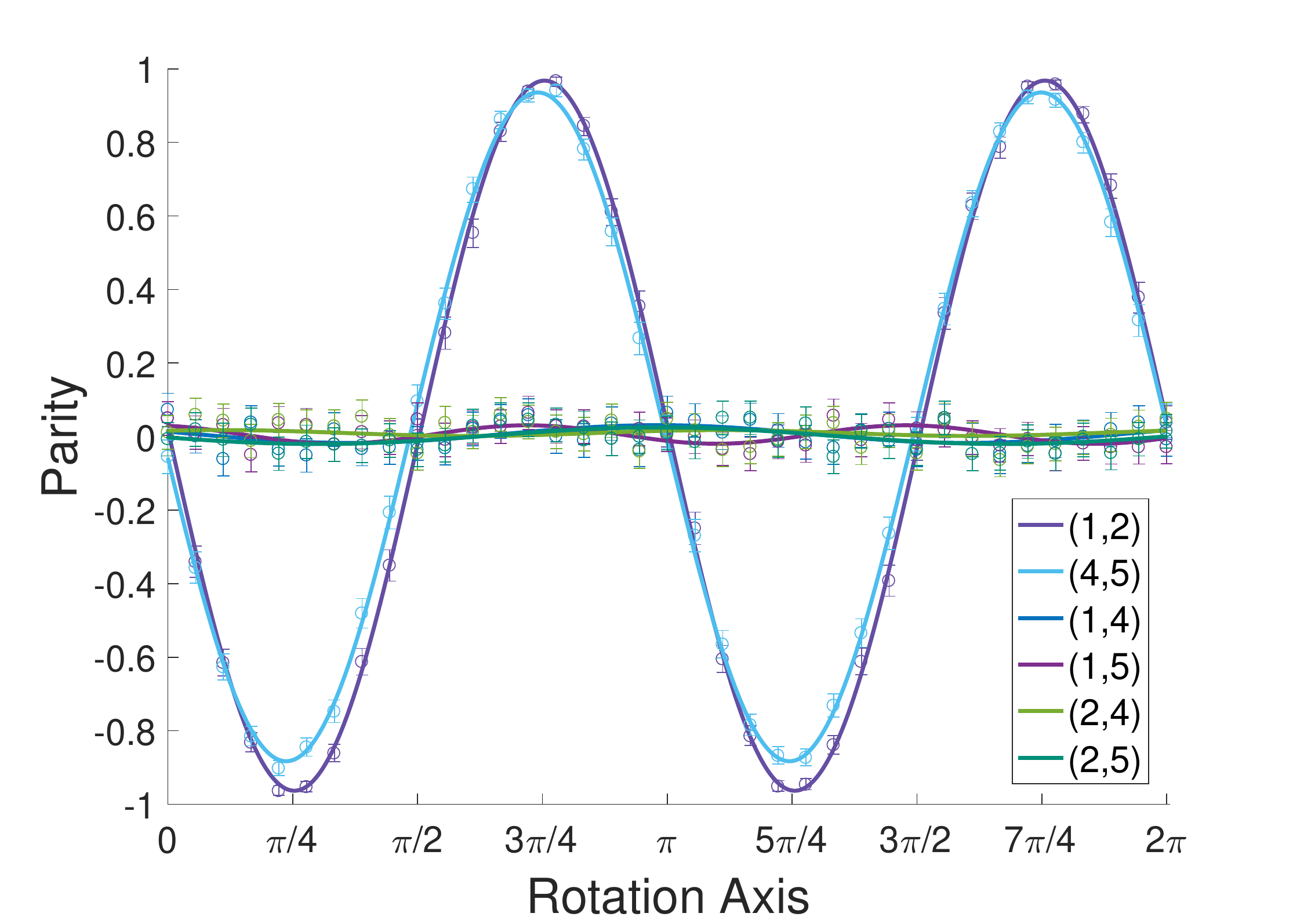}}\\ 
\end{tabular}
\caption[Parity curves for parallel 2-qubit gates on several sets of ions.]{Parity curves and fidelities for parallel $\xxgate$ gates on several sets of ions. Circles indicate data, with matching-color lines indicating calculated fit. 
(a) Ions (1,2) and (3,4), yielding fidelities of 98.4(3)\% and 97.7(3)\% on the respective entangled pairs, with an average crosstalk error of 0.6(3)\%. 
(b) Ions (1,5) and (2,4), yielding fidelities of 96.8(3)\% and 98.1(2)\% on the respective entangled pairs, with an average crosstalk error of 1.7(3)\%. 
(c) Ions (1,3) and (2,5), yielding fidelities of 98.3(3)\% and 97.5(2)\% on the respective entangled pairs, with an average crosstalk error of 0.8(4)\%. 
(d) Ions (1,2) and (4,5), yielding fidelities of 97.2(3)\% and 91.9(3)\% on the respective entangled pairs, with an average crosstalk error of 0.9(3)\%. 
}
\label{fig:MoreParityCurves}
\end{figure*}

Additional parity curves and corresponding gate fidelities are shown in Figure \ref{fig:MoreParityCurves}, with typical fidelities of 96-99\%. 
An exception is the \{(1,2), (4,5)\} gate, for which the (4,5) gate has a fidelity of 91\% (Figure \ref{fig:MoreParityCurves}(d)); however, its phase space closure diagram in \cite{FiggattThesis18} shows that this low fidelity is because the pulse solution found is not ideal.

\afterpage{\clearpage}

\section{Fidelity of Parallel 2-Qubit Entangling Gates with Different Degrees of Entanglement}

\begin{figure}
\centering
\includegraphics[width=\columnwidth,valign=T]{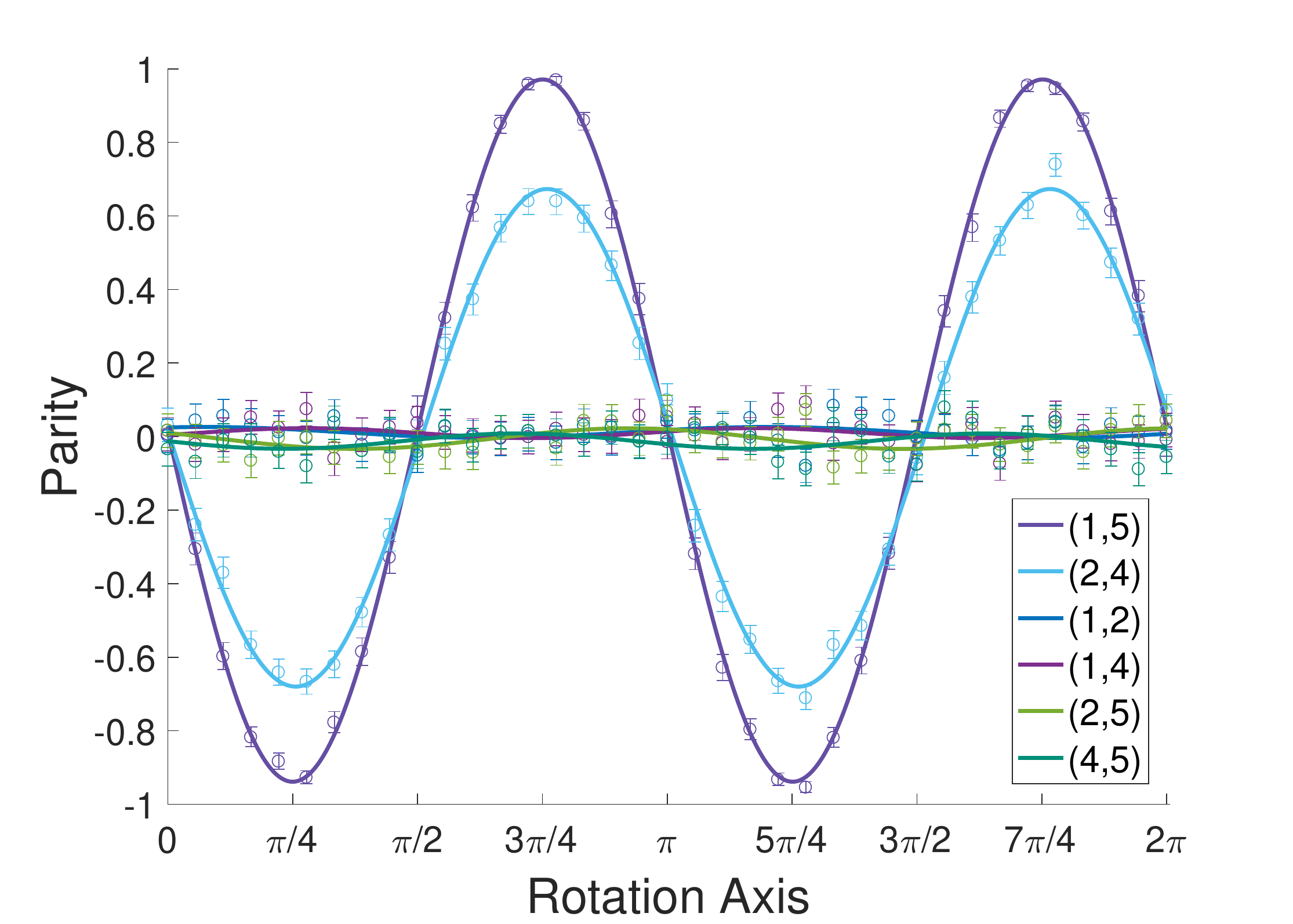}
\caption[Parity curve for parallel 2-qubit gates on (1,5) $\xxgate \left( \frac{\pi}{4} \right)$, (2,4) $\xxgate \left( \frac{\pi}{8} \right)$.]{Parity curve for parallel $\xxgate(\chi)$ gates on ions (1,5) and (2,4), where an $\xxgate \left( \frac{\pi}{4} \right)$ gate is performed on ions (1,5), and an $\xxgate \left( \frac{\pi}{8} \right)$ gate is performed on ions (2,4). Circles indicate data, with matching-color lines indicating calculated fit. This yields fidelities of 96.4(3)\% and 99.4(3)\% on the respective entangled pairs, with an average crosstalk error of 2.2(3)\%.
}
\label{fig:Parity1524pi8}
\end{figure}

Since the $\xxgate$ gates in this parallelization scheme have independent calibration (see next section of Methods), the $\chi$ parameters of the two $\xxgate$ gates are independent. The continuously-variable parameter $\chi$ is directly related to the amount of entanglement generated between the two qubits, given by
\begin{equation}
\xxgate\left(\chi \right)\ket{00} = \frac{1}{\sqrt{2}} \left( \cos\left(\chi \right) \ket{00} - i \sin\left(\chi \right)\ket{11}\right),
\end{equation}
and can be adjusted on the experiment by scaling the power of the overall gate. Consequently, we can simultaneously implement two $\xxgate$ gates with different degrees of entanglement, which may prove useful for some applications. For example, the full adder implementation in the main text requires simultaneously performing an $\xxgate \left( \frac{\pi}{4} \right)$ gate on one pair of qubits, and an $\xxgate \left( \frac{\pi}{8} \right)$ gate on another pair of qubits. To demonstrate this capability, Figure \ref{fig:Parity1524pi8} shows parity scan data for a simultaneous $\xxgate \left( \frac{\pi}{4} \right)$ gate on ions (1,5) and an $\xxgate \left( \frac{\pi}{8} \right)$ gate on ions (2,4). The data is analyzed as in Figures \ref{fig:ParityCurves} and \ref{fig:MoreParityCurves}, but while we use Equation \ref{eqn:FXXpi4} (setting $\chi = \frac{\pi}{4}$) to calculate the fidelity for the (1,5) gate, we use Equation \ref{eqn:FXXChi} and set $\chi = \frac{\pi}{8}$ for the (2,4) gate. The respective gate fidelities are therefore 96.4(3)\% and 99.4(3)\%, with an average crosstalk error of 2.2(3)\%.

\section{Independence of Parallel Gate Calibration}\label{sec:IndependentCalibration}

\begin{figure*}
\centering
\begin{tabular}[c]{c c}
\multicolumn{1}{l}{\bf{(a)}} & \multicolumn{1}{l}{\bf{(b)}} \\
\multicolumn{1}{c}{\includegraphics[width=0.5\textwidth,valign=T]{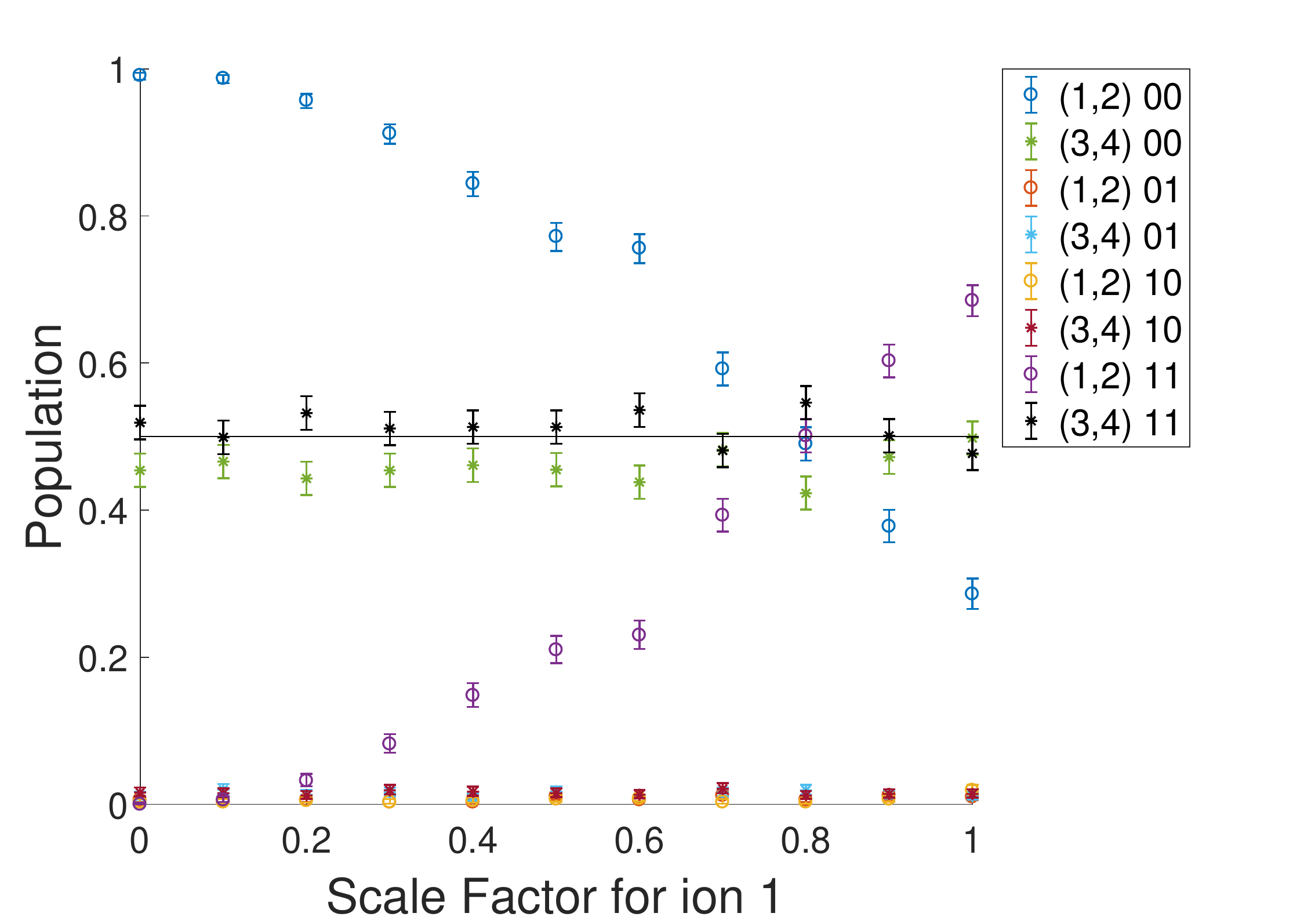}} & \multicolumn{1}{c}{\includegraphics[width=0.5\textwidth,valign=T]{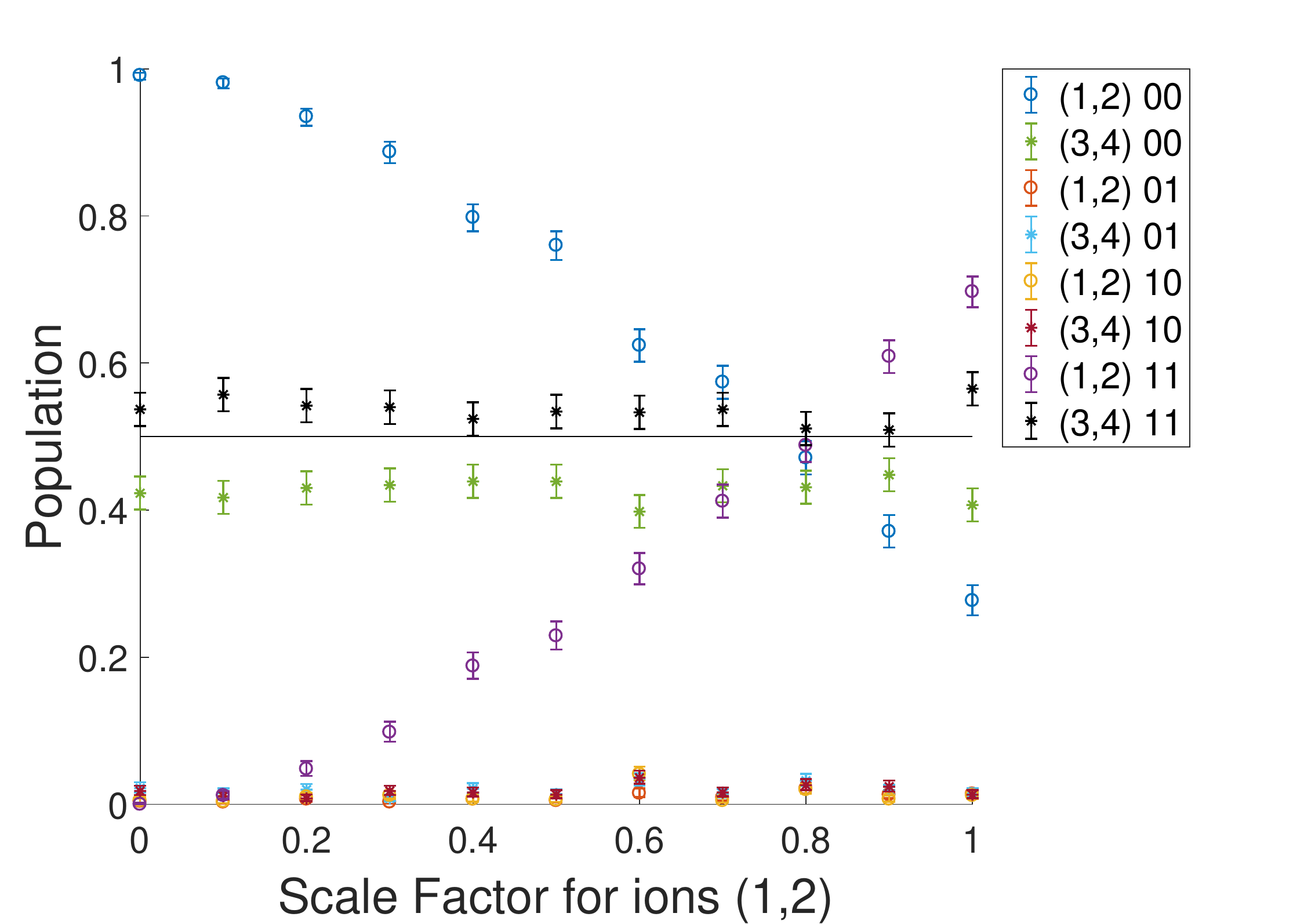}}\\ 
\multicolumn{1}{l}{\bf{(c)}} & \multicolumn{1}{l}{\bf{(d)}}\\
\multicolumn{1}{c}{\includegraphics[width=0.5\textwidth,valign=T]{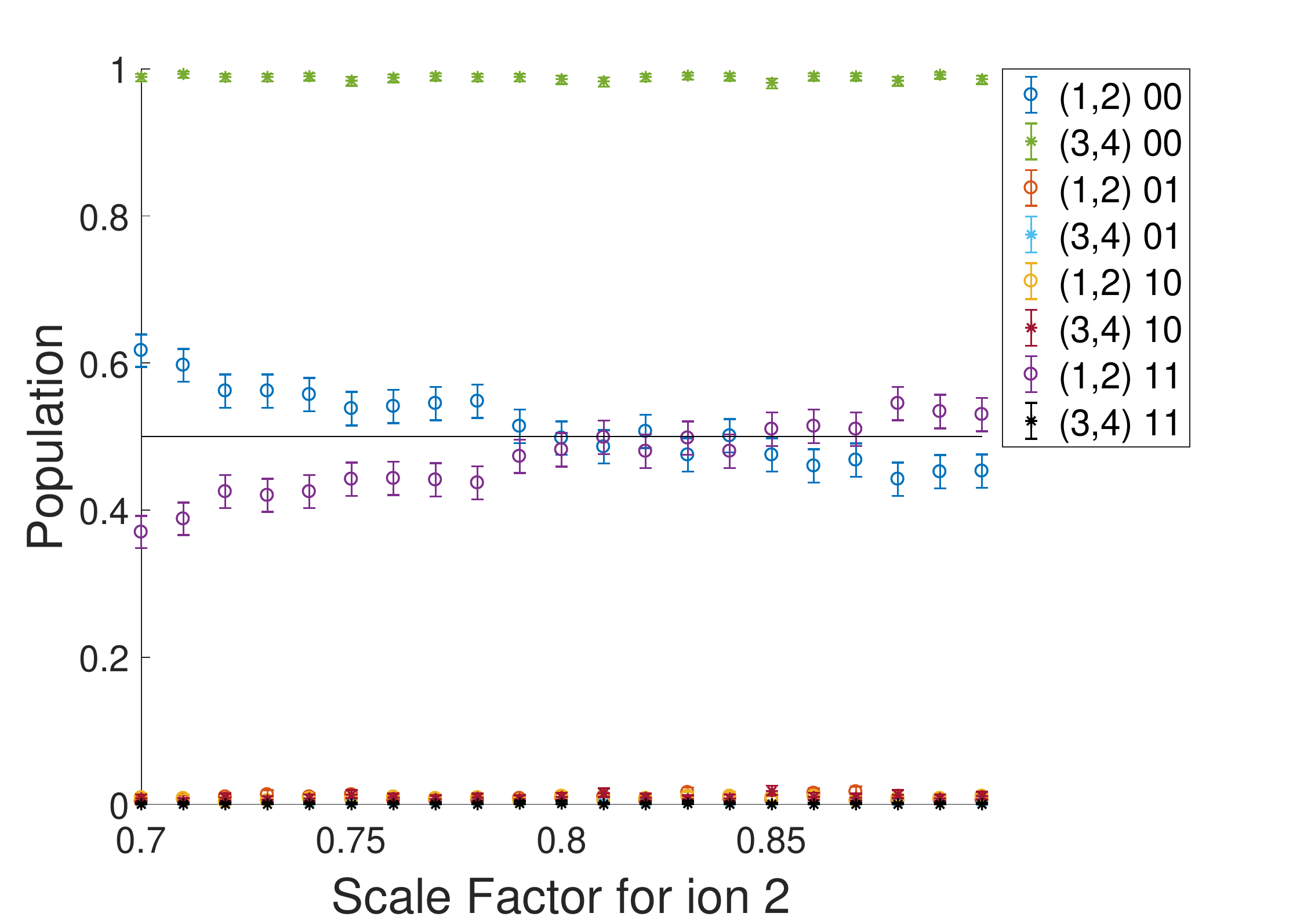}} & \multicolumn{1}{c}{\includegraphics[width=0.5\textwidth,valign=T]{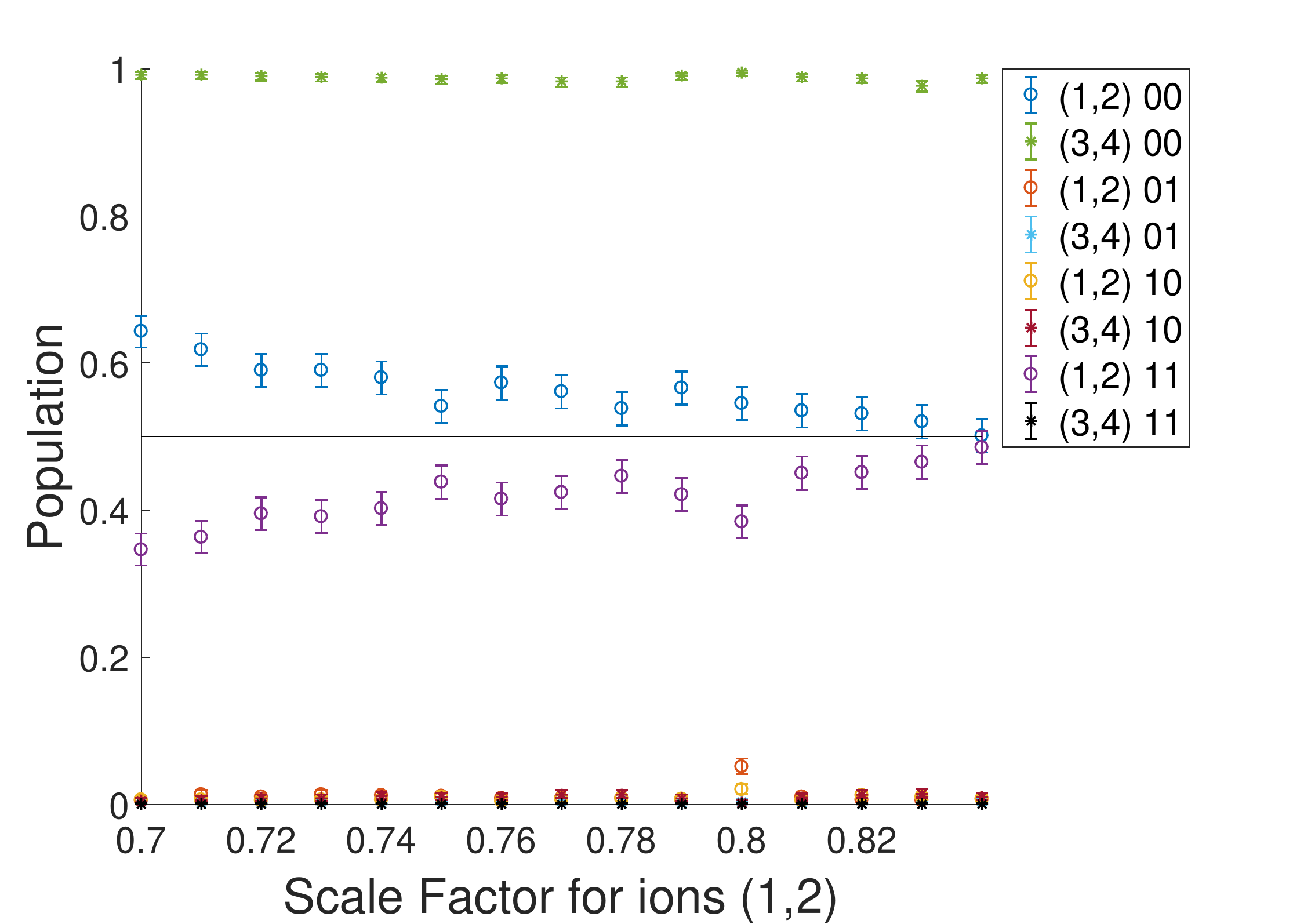}}
\end{tabular}
\caption[Independence of parallel gate calibration.]{Parallel gates can be calibrated independently. (a) Scanning the scale factor on ion 1, with ions (3,4) performing an entangling gate. (b) Scanning the scale factor on ions (1,2), with ions (3,4) performing an entangling gate. (c) Scanning the scale factor on ion 2, with no light on ions (3,4). (d) Scanning the scale factor on ions (1,2), with no light on ions (3,4). }
\label{fig:IndepCalibration}
\end{figure*}

Parallel gates can be calibrated independently from one another by adjusting a scaling factor that controls the overall power on the gate without modifying the pulse shape. Furthermore, adjusting a scaling factor that controls the power on a single ion only affects the gate in which it participates by modifying the total amount of entanglement, without any apparent ill effects on the gate quality. This was confirmed experimentally using parallel operations on ions (1,2) and (3,4) by scanning over the scaling factors associated with ions 1 and 2. Figures \ref{fig:IndepCalibration}(a-b) show two such scans over the scaling factors for ions 1 and 2 while keeping the (3,4) gate ``on'', with the scaling factor for those two ions set near to a fully-entangling gate. Figure \ref{fig:IndepCalibration}(a) shows a scan of just the scaling factor for ion 1 while holding the scale factor for ion 2 constant, and Figure \ref{fig:IndepCalibration}(b) shows a scan over the scaling factor for ions 1 and 2 together. Figures \ref{fig:IndepCalibration}(c-d) show scans over the scaling factors for ions 1 and 2 while keeping the interaction on (3,4) ``off''; the scaling factor for the (3,4) gate is set to 0, so the ions see no light and therefore perform no interaction during the gate. Figure \ref{fig:IndepCalibration}(c) scans the scale factor just on ion 2 while holding the scale on ion 1 constant, and Figure \ref{fig:IndepCalibration}(d) scans the overall scaling factor for ions 1 and 2 together. For all of these scans, as the scaling factors are increased, the population in $\ket{11}$ for ions 1 and 2 increases (and the population in $\ket{00}$ decreases correspondingly), while the $\ket{00}$ and $\ket{11}$ populations for the (3,4) gate remain unchanged.

\section{Optical Power Requirements}

\begin{table}
\begin{center}
\begin{tabular}{ | r || c | c |}
\hline
Parallel Gate Pairs & $R_{||}$, Pair 1 & $R_{||}$, Pair 2 \\ \hline\hline
(1,4) and (2,5) & 4.3 & 1.8 \\ \hline
(1,2) and (3,4) & 7.9 & 5.0 \\ \hline
(1,5) and (2,4) & 2.1 & 1.6 \\ \hline
(1,4) and (2,3) & 4.3 & 3.8 \\ \hline
(1,3) and (2,5) & 0.9 & 1.5 \\ \hline
(1,2) and (4,5) & 2.2 & 2.2 \\
\hline
\end{tabular}
\end{center}
\caption[Comparison of optical power for parallel and single $\xxgate$ gates.]{For each pair of parallel $\xxgate$ gates implemented, we compare the optical power required to perform each component $\xxgate$ with its corresponding stand-alone 2-qubit $\xxgate$ gate by calculating the power ratio $R_{||}$.}
\label{tab:ParallelPower}
\end{table}

While the gate time $\tau_{gate}=250$ $\mu$s for running 2 $\xxgate$ gates in parallel is comparable to that of a single $\xxgate$ gate (and consequently, half the time it would take to execute two $\xxgate$ gates in series), the parallel gates scheme requires more optical power. The Rabi frequency $\Omega$ is proportional to the square root of the beam intensity $I$, $\Omega \propto \sqrt{I_0 I_1}$, where $I_0$ and $I_1$ are the beam intensities for the individual and global beams. We can therefore calculate the ratio $R_{||}$ of the power for a gate executed in parallel to the power required for a single $\xxgate$ gate on the same ions as $R_{||} = \frac{P_{||}}{P_{\xxgate}} = \frac{I_{||}}{I_{\xxgate}} = \left( \frac{\Omega_{||}}{\Omega_{\xxgate}} \right)^2$. 
Intensity is power per unit area, and since the beam sizes do not not vary, the areas cancel out. Measured power ratios for each experimentally implemented gate are shown in Table \ref{tab:ParallelPower}. While some gates required substantially more power (for example, we had some trouble finding a solution for (1,2), (3,4) that was both high-quality and low power), most gates performed in parallel require about two to four times as much power as their singly-performed counterparts. However, a full accounting of power requirements on this experiment must also take into account power wasted by unused beams, and the total time required to perform equivalent operations. Since the individual addressing system has all individual beams on at all times and are dumped after the AOM when not in use (see \cite{Debnath16, FiggattThesis18}), any ion not illuminated corresponds to an individual beam wasting power. Running 2 $\xxgate$ gates in parallel takes $\tau_{gate}=250$ $\mu$s and uses beams each with power $P$ to illuminate 4 ions, but performing those same 2 gates in series using stand-alone $\xxgate$ gates requires time $2\tau_{gate}$ and uses 4 beams each with power $P/4$ to $P/2$ to illuminate 2 ions, wasting 2 beams. This yields a choice of using twice the power (or more) in half the time versus half the power in twice the time; these parallel gates are then very useful when one has more laser power than time.

\section{Optimized Adder Circuit}

\begin{figure*}
\centering
\begin{tabular}[c]{l}

\begin{tabular}[c]{l c  l}
\enspace\enspace
\Qcircuit @C=.75em @R=1.0em @!R {
\lstick{x}		&\gate{R_\text{z}(-\frac{3\pi}{4})} &\gate{R_\text{y}(\frac{\pi}{2})} 	  & \qw				 & \qw						& \multigate{1}{XX(\frac{\pi}{4})} 	& \qw						 & \qw\\
\lstick{y}		&\gate{R_\text{z}(-\frac{\pi}{4})} &\gate{R_\text{y}(\frac{\pi}{2})} 	& \gate{XX(\frac{\pi}{8})} 	&\gate{R_\text{y}(-\frac{\pi}{2})} 	& \ghost{XX(\frac{\pi}{4})}			&\gate{R_\text{z}(-\frac{\pi}{2})}	 & \qw\\
\lstick{C_{in}}	&\gate{R_\text{z}(\frac{\pi}{4})} &\gate{R_\text{y}(\frac{\pi}{2})} 	&  \qw  \qwx \qwx[1] 	 & \qw						& \multigate{1}{XX(\frac{\pi}{8})} 	&\gate{R_\text{z}(-\frac{\pi}{2})}	 & \qw	\\
\lstick{0} 		&\gate{R_\text{x}(-\frac{\pi}{2})} & \qw						& \gate{XX(\frac{\pi}{8})}	 & \qw						& \ghost{XX(\frac{\pi}{8})}			& \qw						& \qw	\gategroup{1}{6}{4}{6}{.7em}{--}
}
& &
\\ 
\\
\\
\enspace\enspace
\Qcircuit @C=.75em @R=1.0em @!R {
\lstick{\cdots} 	& \gate{XX(\frac{\pi}{8})}	& \qw&\gate{R_\text{y}(-\frac{\pi}{2})}	& \qw			& \qw & \qw & \qw  & \push{x}\\
\lstick{\cdots}	&  \qw  \qwx \qwx[1]		&\multigate{1}{XX(\frac{\pi}{4})}&\gate{R_\text{y}(\frac{\pi}{2})}	& \qw			&\gate{R_\text{z}(\pi)}	& \qw & \qw & \push{y'}\\
\lstick{\cdots}	&  \qw  \qwx \qwx[1]		&\ghost{XX(\frac{\pi}{4})}&\gate{R_\text{y}(\frac{\pi}{2})}	&\multigate{1}{XX(-\frac{\pi}{8})}	&\gate{R_\text{y}(-\frac{\pi}{2})}	&\gate{R_\text{z}(\frac{\pi}{4})}		& \qw & \push{S}\\
\lstick{\cdots}	& \gate{XX(\frac{\pi}{8})}	&  \push{\rule{4.2em}{0.04em}}\qw	& \qw						&\ghost{XX(-\frac{\pi}{8})}	& \qw & \qw & \qw & \push{C_{out}}  \gategroup{1}{2}{4}{3}{.7em}{--}
}

\end{tabular}

\end{tabular}
\caption[Full adder circuits.]{Application-optimized full adder implementation using $\xxgate(\chi)$, $\rgate_\text{x}(\theta)$, and $\rgate_\text{y}(\theta)$ gates. The two parallel 2-qubit operations are outlined in dashed boxes.}
\label{fig:OptimizedAdderCircuit}
\end{figure*}
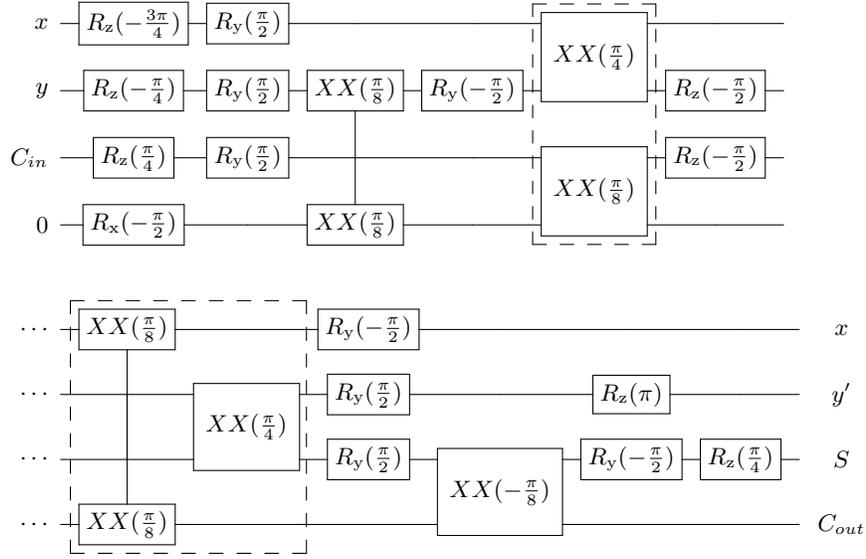

The optimized full adder circuit to be implemented on the experiment, shown in Figure \ref{fig:OptimizedAdderCircuit}, is constructed from Figure \ref{fig:AdderCircuits}(b) by combining the $\cnotgate$, $\textsc{C}(\textsc{V})$, and $\textsc{C}(\textsc{V}^{\dag})$ gates from Figure 5.12 of \cite{FiggattThesis18} and further optimizing the rotations per the method described in Section 5.2.1 of \cite{FiggattThesis18}. 
The two parallel 2-qubit operations are outlined in dashed boxes.

\afterpage{\clearpage}
\newpage
\pagebreak

\twocolumngrid

\afterpage{\clearpage}

\end{document}